\documentclass[11pt]{amsart}
\pdfoutput=1
\usepackage{amssymb}
\usepackage{epsfig}
\usepackage{comment}
\usepackage{esint}
\usepackage{algorithm}
\usepackage{algorithmic}
\usepackage{subcaption}
\usepackage{units}

\theoremstyle{definition}

\theoremstyle{remark}

\graphicspath {
    {./}
    {./Figures/}
}

\begin{document}

\title[Oncotripsy]{A Dynamical Model of Oncotripsy by Mechanical Cell Fatigue: Selective Cancer Cell Ablation by Low-Intensity Pulsed Ultrasound (LIPUS)}

\author[E.~F.~Schibber {\sl et al.}]{E.~F.~Schibber$^{1}$, D.~Mittelstein$^{1}$, M.~Gharib$^{1}$, M.~Shapiro$^{1}$, P.~Lee$^{2}$ and M.~Ortiz$^{1}$}

\address{$^{1}$Division of Engineering and Applied Science, California Institute of Technology, 1200 East California Boulevard, Pasadena, CA 91125, USA.\\
$^{2}$ Department of Immuno-Oncology, City of Hope National Medical Center, 1500 E Duarte Rd, Duarte, CA 91010, USA}

\keywords{oncotripsy, ultrasound, LIPUS, biomechanics, fatigue}

%\corres{M.~Ortiz\\
\email{ortiz@caltech.edu}%}

\begin{abstract}
The method of {\sl oncotripsy}, first proposed in \cite{Heyden2016}, exploits aberrations in the material properties and morphology of cancerous cells in order to ablate them selectively by means of tuned low-intensity pulsed ultrasound (LIPUS). We propose a dynamical model of oncotripsy that follows as an application of cell dynamics, statistical mechanical theory of network elasticity and 'birth-death' kinetics to describe processes of damage and repair of the cytoskeleton. We also develop a reduced dynamical model that approximates the three-dimensional dynamics of the cell and facilitates parametric studies, including sensitivity analysis and process optimization. We show that the dynamical model predicts---and provides a conceptual basis for understanding---the oncotripsy effect and other trends in the data of Mittelstein {\sl et al.} \cite{Mittelstein:2019} for cells in suspension, including the dependence of cell-death curves on cell and process parameters.\end{abstract}

\maketitle

% \tableofcontents

%%%%%%%%%%%%%%%%%%%%%%%
\section{Introduction}
%%%%%%%%%%%%%%%%%%%%%%%

\begin{figure}
	\centering
	\includegraphics[width=0.75\textwidth]{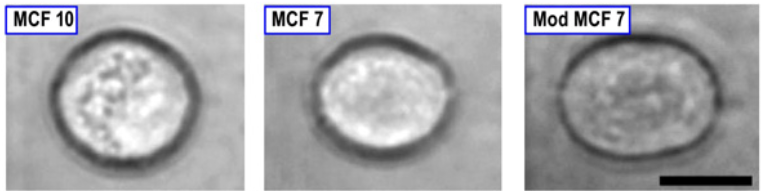}
    \caption{Optical images showing deformability on three breast cells due to a constant stretching laser power of 600mW. Deformability increases in the cancerous MCF-7 and ModMCF-7 cells in comparison to the healthy cell MCF-10. Reproduced from \cite{Suresh2007a} and \cite{Guck:2005}.}	
	\label{fig:Suresh2007Fig12}
\end{figure}

\begin{figure}
	\centering
	\includegraphics[width=0.75\textwidth]{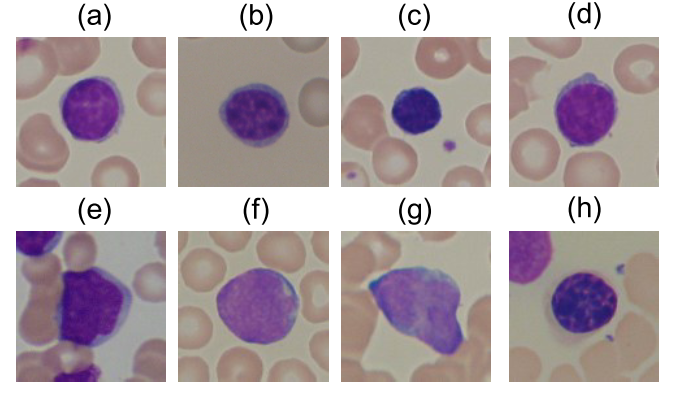}
    \caption{(a-d) Healthy lymphocyte cells from non-Acute Lymphoblastic Leukemia patients. (e-h) Probable lymphoblast cells showing marked differences in size and morphology with respect to the healthy cells. Reproduced from~\cite{Labati2011}.}	
	\label{fig:Labati2011Fig6}
\end{figure}

The method of {\sl oncotripsy}, first proposed in \cite{Heyden2016}, exploits aberrations in the material properties and morphology of cancerous cells, cf., e.~g., Figs.~\ref{fig:Suresh2007Fig12} and \ref{fig:Labati2011Fig6}, in order to ablate them selectively by means of tuned low-intensity ultrasound. A wealth of observational evidence reveals that a substantial size difference between normal nuclei, with an average diameter of $7$ to $9$ microns, and malignant nuclei, which can reach a diameter of over $50$ microns, often characterizes malignancy \cite{Berman:2011}. Using atomic force microscopy, \cite{Cross:2007} reported the stiffness of live metastatic cancer cells taken from pleural fluids of patients with suspected lung, breast and pancreas cancer. They found that the cell stiffness of metastatic cancer cells is more than 70\% softer than the benign cells that line the body cavity. Swaminathan {\sl et al.} \cite{Swaminathan:2011} applied a magnetic tweezer system to measure that stiffness of human ovarian cancer cell lines and found that cells with the highest invasion and migratory potential are up to five times softer than healthy cells \cite{Swaminathan:2011}. Experimental investigations of hepatocellular carcinoma cells (HCC) have also found that an increase in stiffness of the extracellular matrix (ECM) promotes HCC proliferation \cite{Schrader:2010} and advances malignant growth \cite{Levental:2009}.

Owing to these and other similar observed aberrations in material properties and morphology attendant to malignancy, the eigenfrequencies at which cell resonance occurs are expected to differ markedly between healthy and cancerous cells. In a recent numerical study \cite{Heyden2016}, Heyden and Ortiz have shown that HCC natural frequencies lie above those of healthy cells, with a typical gap in the lowest natural frequency of about $37$\,kHz. For instance, they computed the fundamental frequency of HCC to be of the order 80 kHz and of the order of $43$ kHz for the healthy cells. Heyden and Ortiz \cite{Heyden2016} posited that, by exploiting this {\sl spectral gap}, cancerous cells can be selectively ablated by means of carefully tuned ultrasound while simultaneously leaving normal cells intact, an effect that they referred to as {\sl oncotripsy}. Specifically, by studying numerically the vibrational response of HCC and healthy cells, Heyden and Ortiz \cite{Heyden2016} found that, by carefully tuning the frequency of the harmonic excitation, lysis of the HCC nucleolus membrane could be induced selectively at no risk to healthy cells. They also estimated the acoustic density required for oncotripsy to operate to be in the Low-Intensity Ultrasound (LIPUS) range. This low-intensity requirement sets oncotripsy apart from High-Intensity Focused Ultrasound (HIFU), which acts via thermal ablation and is non-specific, with no selectivity for cancer cells.

The first numerical calculations of Heyden and Ortiz \cite{Heyden2016} neglected viscoelasticity and damping in the cell and ECM. Under these conditions, the resonant response of the cells exhibits rapid linear growth in time and the cells are predicted to attain lysis relatively quickly. However, experimental studies suggest that the material behavior of the different cell constituents is viscoelastic \cite{Panorchan:2006, Guilak:2000, Zhang:2002, Janmey:1998}. In a subsequent study \cite{Heyden:2016b}, Heyden and Ortiz investigated the influence of viscoelasticity on the oncotripsy effect. They assumed Rayleigh damping and estimated the damping coefficients from dynamic atomic force microscopy (AFM) experiments on live fibroblast cells in buffer solutions \cite{Cartagena:2014}. They concluded that, for these cells, the main effect of viscoelasticity is a modest reduction in the resonant natural frequencies of the cells and an equally modest increase of the time to lysis of the cancerous cells. On the basis of these results, they speculated that oncotripsy remains viable when viscoelasticity is taken into account.

Following these leads, Mittelstein {\sl et al.} \cite{Mittelstein:2019} have endeavored to assess the oncotripsy effect in carefully designed laboratory tests involving a number of cancerous cell lines in aqueous suspension. They have developed a system for testing oncoptripsy that includes a tunable source of ultrasonic transduction in signal communication with a control system that allows control of several parameters, including frequency and pulse duration, of the ultrasonic transduction. Transducers were selected to produce ultrasound pulses in the frequency range of approximately 100 kHz to 1 MHz, a pulse duration range of 1 ms to 1 s, acoustic intensity up to 5 W/cm${}^2$, and output pressure up to 2 MPa. The instrumentation of the system allows the measurement of estimated cell death rates as a function of frequency, pressure, pulse duration, duty cycle and number of cycles.

In agreement with the original oncotripsy concept, the experiments confirm that the application of low-intensity pulsed ultrasound (LIPUS) can indeed result in high death rates in the cancerous cell population {\sl selectively}, i.~e.,  simultaneously with small or zero death rates among healthy cells. The death and survival rates depend critically on the frequency of the ultrasound, indicative of a dynamical response of the cells. The oncotripsy effect is maximum at a certain frequency, and diminishes at both larger and smaller frequencies, also indicative of a resonant response of the cells. However, under the conditions of the experiments, cell death is observed to require the application of a much larger number of ultrasound cycles than anticipated by either \cite{Heyden2016} or \cite{Heyden:2016b}, suggesting that the dynamics of cells in aqueous suspension is much more heavily damped than estimated in \cite{Heyden:2016b} based on the AFM measurements of \cite{Cartagena:2014}. The observations reported by Mittelstein {\sl et al.} \cite{Mittelstein:2019} suggest that, under the conditions of the experiment, cell death occurs though a process of slow accumulation of damage over many cycles, instead of the rapid rupture of one of the cell membranes, as hypothesized in \cite{Heyden2016}.

\begin{figure*}[ht!]
    \centering
    \includegraphics[width=0.7\textwidth]{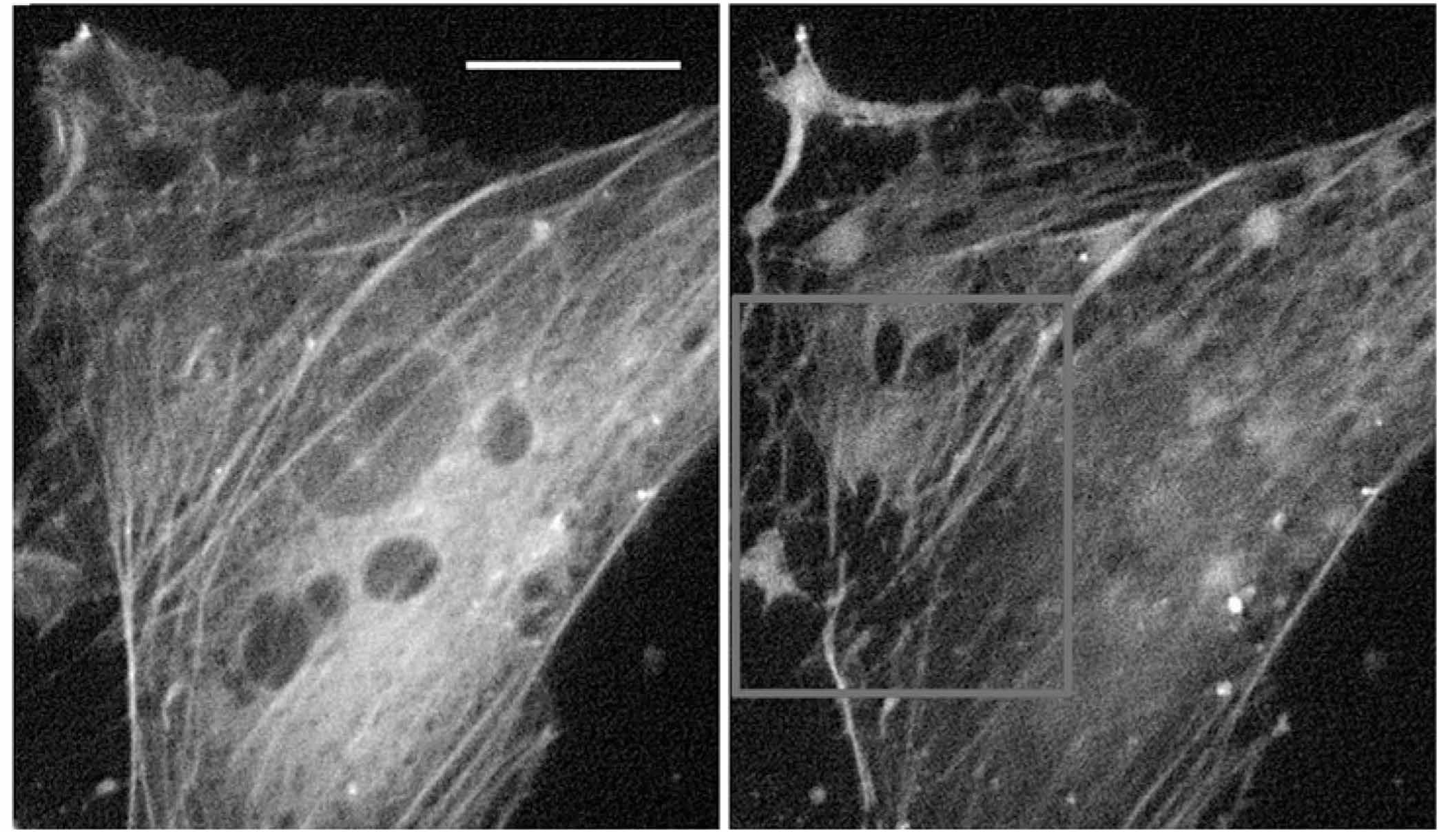}
    \caption{\small Live yellow fluorescent protein (YFP) tagged actin network staining of cells before and 5 min after exposure to 290 kPa acoustic pressure showing massive fiber disruption (reproduced from \cite{Mizrahi:2012}). Scale bar 10 $\mu$m.} \label{gGZ1YS}
\end{figure*}

A number of experimental investigations suggest a mechanistic basis for the oncotripsy effect. The susceptibility of the cytoskeleton dynamics to therapeutic ultrasound, at strains of the order of $10^{-5}$ and frequencies in the MHz range, has been noted by \cite{Mizrahi:2012}. At low acoustic intensities, no structural network changes are observed over the duration of the experiments. By contrast, at sufficiently high acoustic intensities the actin network is progressively disrupted and disassembles within three minutes following exposure, Fig.~\ref{gGZ1YS}. This disruption is accompanied by a 50\% reduction in cell stiffness. Remarkably, after exposure to moderate acoustic intensities the stiffness of the cell gradually recovers and returns to its initial value. The mechanisms of actin stress-fiber repair have been extensively studied and are reasonably well-understood at present, cf., e.~g., \cite{Smith:2014, Nakamura:2018} and references therein. By contrast, at high acoustic intensities no recovery takes place after cessation over the span of observation.

\begin{figure*}[ht!]
    \centering
    \includegraphics[width=0.7\textwidth]{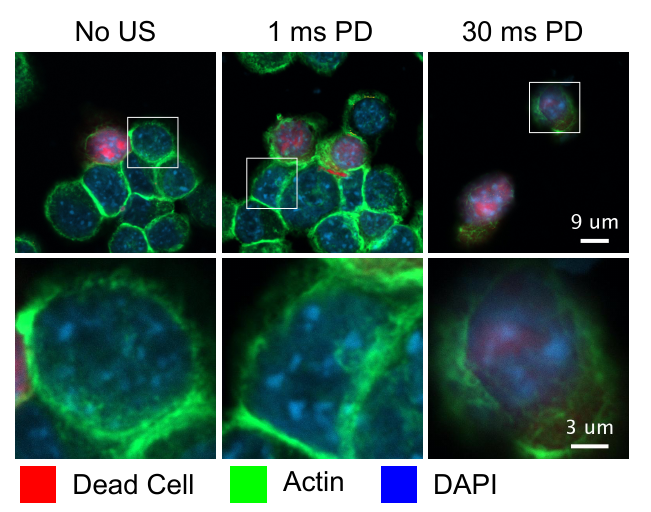}
    \caption{\small Confocal microscopy of CT-26 cells immediately after LIPUS treatment at 500kHz, focal pressure of 1.4MPa and pulse durations 0 ms (control), 1 ms and 30 ms (reproduced from \cite{Mittelstein:2019}). Dead cells stained red with fixable LIVE/DEAD, actin cytoskeleton stained green using phalloiding, and nucleus stained blue with DAPI. Confocal images shows disrupted actin cytoskeleton ring and significantly decreased actin stain intensity. Microscopy suggests LIPUS cytodisruption is coupled with persistent cytoskeletal disruption.} \label{x6sPlV}
\end{figure*}

To gain insight into the biomolecular mechanisms of LIPUS cytodisruption, Mittelstein {\sl et al.} \cite{Mittelstein:2019} examined CT-26 cells after $2$-minute LIPUS treatment at $500$ kHz and focal pressure of $1.4$ MPa. To evaluate the effect of LIPUS on the cytoskeleton, they plated CT-26 cells after LIPUS and performed confocal microscopy immediately after insonation. Confocal images show the actin cytoskeleton, stained with phalloidin-conjugated green dye, as a ring on the cell periphery, Fig.~\ref{x6sPlV}. This ring is disrupted and shows diminished fluorescence for a 30 ms pulse duration, suggesting that cytodisruption is coupled with persistent cytoskeleton disruption. These observations are consistent with reports for other systems that LIPUS disrupts the cellular cytoskeleton \cite{Noriega:2013, Samandari:2017}. In contrast, with $1$ ms pulse duration, the actin cytoskeleton appears unchanged from the negative control. Mittelstein {\sl et al.} \cite{Mittelstein:2019} conclude that these observations suggest that LIPUS induces actin cytoskeletal disruption and activates apoptotic cell-death pathways.

In the present work, we argue that these competing mechanisms of cytoskeletal disruption and self-repair, when coupled to the---possibly resonant---dynamics of the cells over many insonation cycles, underlie the oncotripsy observations of Mittelstein {\sl et al.} \cite{Mittelstein:2019}. Based on this hypothesis, we develop a plausible theoretical model of oncotripsy that accounts for several of the key experimental observations of Mittelstein {\sl et al.} \cite{Mittelstein:2019}, including the dependence of the cell death rates on frequency, pulsing characteristics and number of cycles. We posit that, under the conditions of the experiments, cells in suspension subjected to LIPUS act as frequency-dependent resonators and that the evolution of the cells is the result of competing mechanisms of high-cycle cumulative damage and healing of the cytoskeleton. We recall that structural materials can fail at load levels well below their static strength through processes of slow incremental accumulation of damage when subjected to a large number (millions) of loading cycles, a phenomenon known as {\sl mechanical fatigue} \cite{Suresh:1998}. Likewise, whereas one single LIPUS pulse is unlikely to cause significant cytoskeletal damage, we posit that over millions of cycles damage can accumulate to levels that render the cell unviable and cause it to die. By analogy to structural materials, we refer to the hypothesized necrosis mechanism as {\sl mechanical cell fatigue}.

We note that, whereas the elasticity, rheology and remodeling of the cytoskeleton have been extensively studied in the past (cf., e.~g., \cite{Lim2006, Mofrad:2009, Fletcher2010, Jensen:2015} and references therein), no model of cumulative damage and mechanical cell fatigue appears to have been as yet proposed. The model proposed in this work follows as an application of cell dynamics, statistical mechanical theory of network elasticity and 'birth-death' kinetics to describe processes of damage and repair of the cytoskeleton. We also develop a reduced dynamical model that approximates the three-dimensional dynamics of the cell and facilitates parametric studies, including sensitivity analysis and process optimization. The reduced dynamical system encompasses the relative motion of the nucleus with respect to the cell membrane and a state variable measuring the extent of damage to the cytoskeleton. The cell membrane is assumed to move rigidly according to the particle velocity induced in the water by the insonation. The dynamical system evolves in time as a result of structural dynamics and kinetics of cytoskeletal damage and repair. The resulting dynamics is complex and exhibits behavior on multiple time scales, including the period of vibration and attenuation, the characteristic time of cytoskeletal healing, the pulsing period and the time of exposure to the ultrasound. We show that this multi-time scale response can effectively be accounted for by recourse to WKB asymptotics and methods of weak convergence \cite{Bender:1978}. We also account for cell variability and estimate the attendant variance of the time-to-death of a cell population using simple linear sensitivity analysis. The reduced dynamical model predicts, analytically up to quadratures, the response of a cell population to LIPUS as a function of fundamental cell properties and process parameters. We show, by way of partial validation, that the reduced dynamical model indeed predicts---and provides a conceptual basis for understanding---the oncotripsy effect and other trends in the data of Mittelstein {\sl et al.} \cite{Mittelstein:2019}, including the dependence of cell-death curves on pulse duration and duty cycle.

\section{Experimental basis}\label{i9UQnX}

We begin with a brief summary of the experimental system developed by Mittelstein {\sl et al.} \cite{Mittelstein:2019}, as well as data and observations resulting from the study that are directly relevant to the present work. Their original publication maybe consulted for a complete account.

\begin{figure}
	\centering
	\begin{subfigure}[]{0.45\textwidth}
		\centering
		\includegraphics[width = 0.99\textwidth]{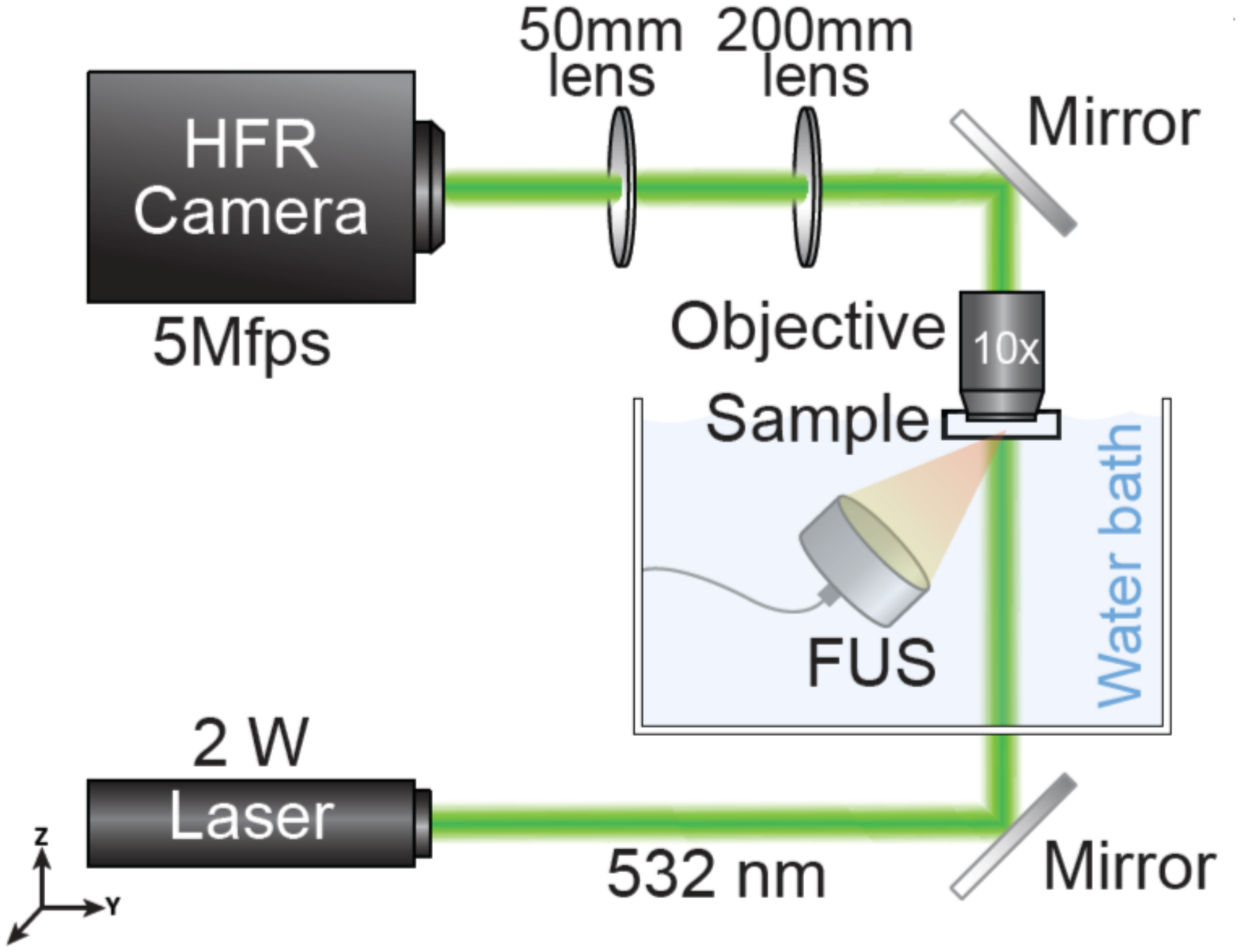}
		\caption[]{}
	\end{subfigure}
	\begin{subfigure}[]{0.45\textwidth}
		\centering
		\includegraphics[width = 0.99\textwidth]{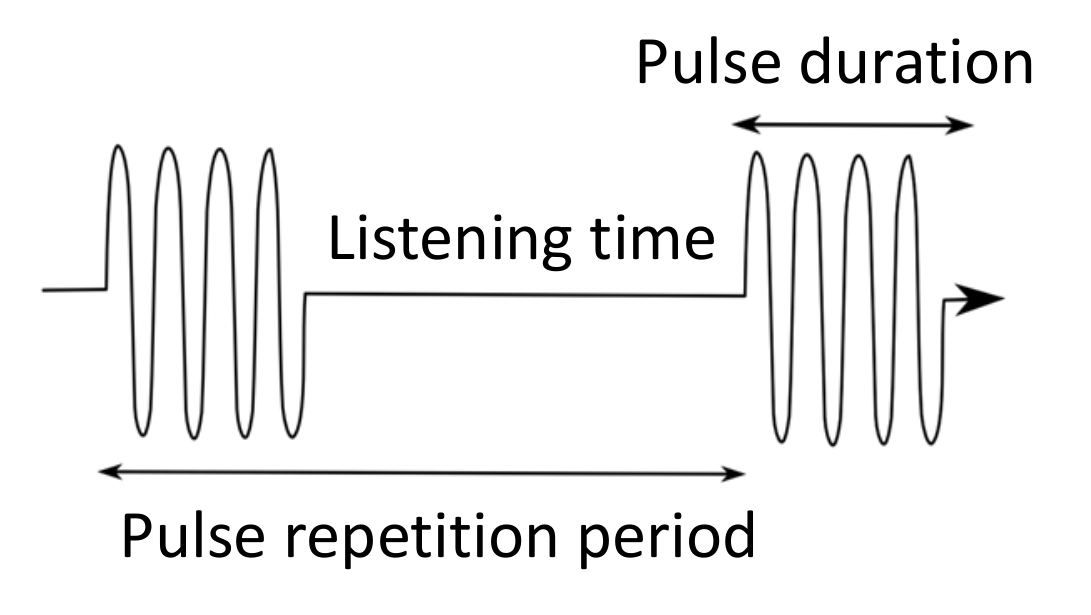}
		\caption[]{}
	\end{subfigure}
    \caption{\small Experimental setup of Mittelstein {\sl et al.} \cite{Mittelstein:2019}. a) Schematic drawing of the LIPUS system and high frame-rate camera setup enabling cellular imaging at a frame rate of 5 MHz. b) Schematic of pulse duty cycle.} \label{8w5RPa}
\end{figure}

\subsection{Experimental system}

The experimental setup, Fig.~\ref{8w5RPa}a, was developed to investigate the response of cells in aqueous suspension to ultrasound insonation \cite{Mittelstein:2019}. Suspension cells are placed with a mylar film pocket that is submerged within a water bath. The cells within the pocket are thus in acoustic contact with the ultrasound transducer.  The investigation indicated that the cell-disruption effect through low intensity pulsed ultrasound (LIPUS) requires the presence of spatial standing waves, which are generated by the reflection of the ultrasound wave off of an acrylic or metal acoustic reflector. Several hypotheses for the requirement of a standing wave are explored in \cite{Mittelstein:2019}. The transducer in the water tank is positioned directly incident with the mylar pocket such that the acoustic axis is perpendicular with the optical axis which is illuminated by laser light. The mylar pocket is supported by a three-sided acrylic frame. One side of this frame serves as an acoustic reflector to form the standing waves. A water immersion pan-fluor objective is lowered into water bath and a series of prism mirrors and converging lenses deliver the image into a high-speed camera. Images are acquired 100 ms after the arrival time of the pulse to observe the effect of prolonged ultrasound exposure.

The experiments aimed to isolate the mechanical effects of ultrasound by preventing local heating from taking place.  In order to maintain low intensity ultrasound conditions (Ispta $<$ 5 W/cm2), pulsed ultrasound was performed as shown in Fig.~\ref{8w5RPa}b. LIPUS was applied at a 10\% duty cycle. However the pulsing parameters were varied in order to investigate their role on ultrasound cytodisruption.  The pulse duration corresponds to the length of each pulse during which the ultrasound is on.  By varying the pulse duration, while maintaining a constant duty cycle, the pulsing pattern of the ultrasound applied to the cells can be modified while maintaining constant acoustic energy deposited on the cells.  To further investigate the effects of modifying ultrasound parameters on cytodisruption, three different transducers operating at 300 kHz, 500 kHz, and 670 kHz were used during this investigation.  To provide consistent comparisons, they were configured to produce a peak negative pressure of 1.4 MPa at their focus in free water.

\begin{figure}[h]
	\centering
	\begin{subfigure}[b]{0.375\textwidth}
		\includegraphics[width=0.99\textwidth]{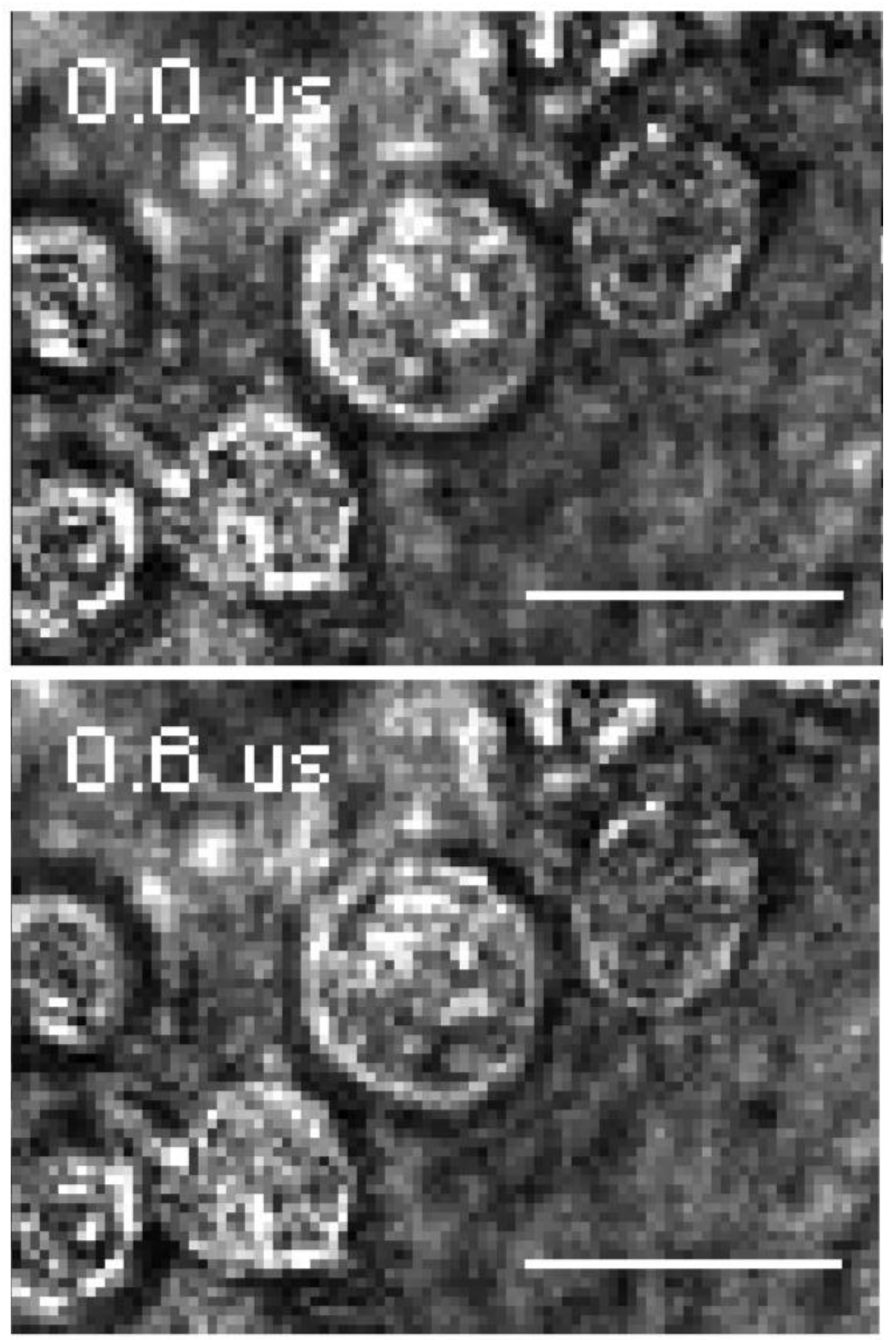}
		\caption{}
	\end{subfigure}
    $\quad$
	\begin{subfigure}[b]{0.525\textwidth}
		\includegraphics[width=0.99\textwidth]{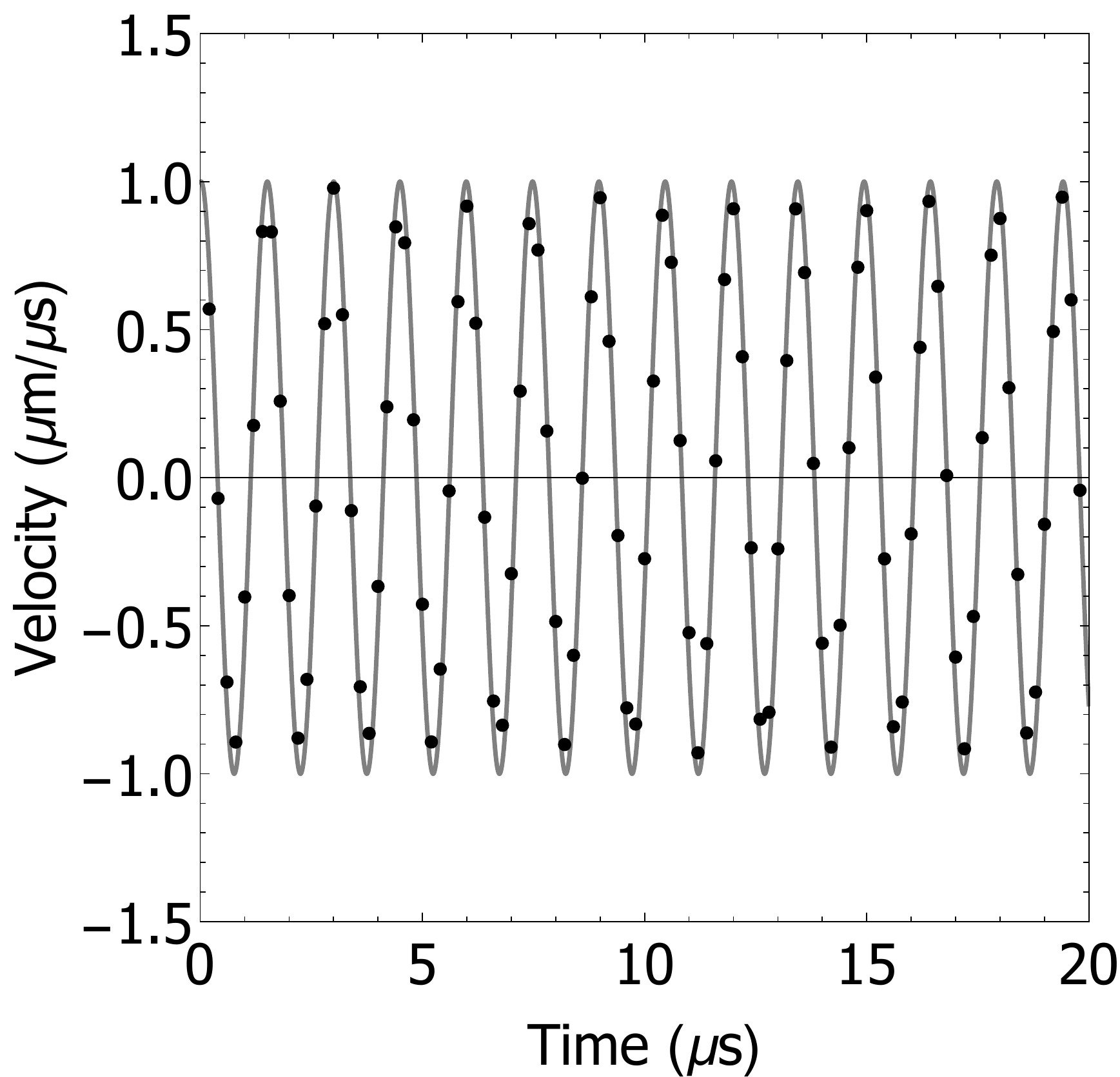}
		\caption{}
	\end{subfigure}
    \caption{a) Frames from video captured by Mittelstein {\sl et al.} \cite{Mittelstein:2019} and processed with Ncorr\cite{Blaber2015}  (scale bar 10 microns). b) Measured velocity of K-562 cell under an incident plane wave of focal pressure amplitude $P_0=1.4$ MPa and excitation frequency $f_0=670$ kHz. }\label{Bk9qt4}
\end{figure}

\subsection{Cell motion}
The recordings show that the entire field-of-view oscillates in the direction of ultrasound propagation with minimal observable cell membrane deformation, Fig.~\ref{Bk9qt4}a. The damping out of cell membrane oscillations is expected given the exceedingly low Reynolds number characteristic of the cell dynamics in aqueous suspension. Fig.~\ref{Bk9qt4}b shows the measured trajectory of a K-562 cell upon insonation of focal pressure of $P_0=1.4$ MPa, frequency $f_0$=670 kHz and wavelength $\lambda = 2.2$ mm. As may be seen from the figure, the cell undergoes an ostensibly harmonic motion. The period of the motion is $T =1.4$ $\mu$s, which corresponds to a frequency of $f=714$ kHz. In addition, the amplitudes of the motion in the $x$- and $y$-directions are $u_x = 0.23$ $\mu$m and $u_y = 0.022$ $\mu$m, respectively, for a total displacement amplitude of $u = \sqrt{u_x^2 + u_y^2} = 0.231$ $\mu$m and a velocity amplitude of $v = 2 \pi f u = 1.037$ m/s. By way of reference, the particle velocity amplitude of the medium is $v_0 = P_0/\rho_0 c_0 = 0.97$ m/s, where $\rho_0 = 1000$ Kg/m${}^3$ is the mass density of water and $c_0 = 1450$ m/s is its sound speed. We thus conclude that, as expected for the long wavelength of the insonation relative to the cell size, the cells move ostensibly at the particle velocity of the fluid.

\begin{table}\small
\begin{center}
\begin{tabular}{p{0.5in}p{1.0in}p{1.0in}p{1.0in}p{1.0in}}
	\hline
	\textbf{Cell Line}& \textbf{Morphology} & \textbf{Tissue}  & \textbf{Disease} & \textbf{Source} \\
    \hline
	K-562	   & Lymphoblast	& Lymphocyte    & Chronic myeologeneous leukemia & Human cell line\\
	U-937	   & Monocyte       & Lymphocyte   & Pleura/pleural effusion, lymphocyte, myeloid & Human cell line\\
	T-Cells & Lymphocyte & Peripheral blood cells, isolated CD3+ & & Human primary cells \\
    \hline\bigskip
\end{tabular}
\caption{Haematopoietic and lymphoid malignancies tumor cells used in the experiments of Mittelstein {\sl et al.} \cite{Mittelstein:2019} classified by morphology, type and disease.}
\label{tbl:varprop}
\end{center}
\end{table}	

\begin{figure}[h]
	
	\centering
	\includegraphics[width=1\textwidth,trim={1cm 5cm 0cm 10cm},clip]{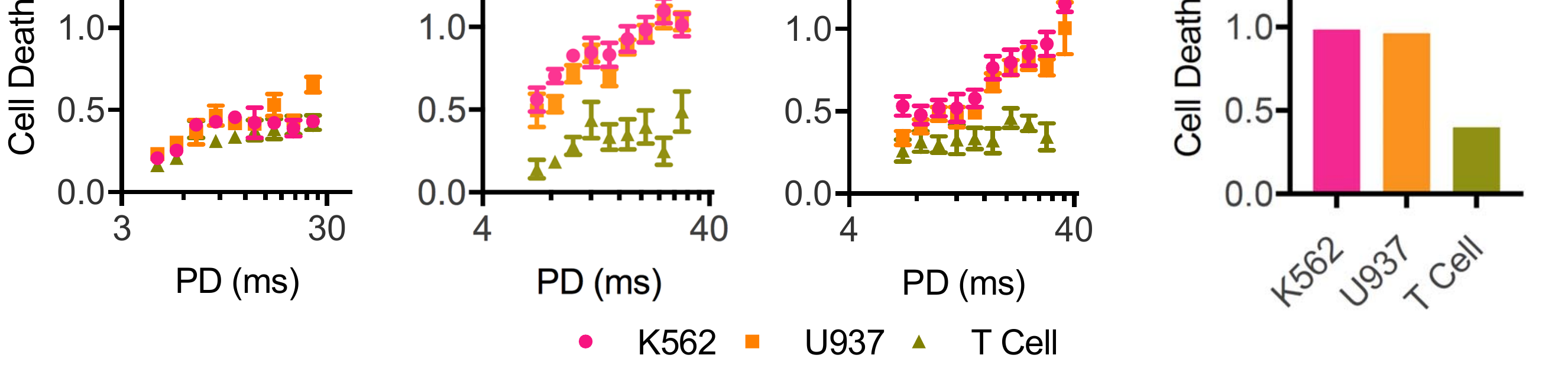}
	\begin{subfigure}[b]{0.75\textwidth}
	\caption{}
\end{subfigure}
	\begin{subfigure}[b]{0.001\textwidth}
	\caption{}
\end{subfigure}
    \caption{Tests of cancerous K562 and U937 cells and healthy CD4 T-cells at a PNP of 0.7MPa and a time of exposure of 60 seconds, showing the effect of frequency and pulse duration on cell death rates. In call cases, the pulse duration is $10\%$ of the total pulse repetition period. (a) Cell death fraction vs pulse duration, and (b) cell death fraction at 20 ms pulse duration vs type. Reproduced from \cite{Mittelstein:2019}.}\label{v0EZTl}
\end{figure}

\begin{figure}[h]
	\centering
	\begin{subfigure}[b]{0.45\textwidth}
		\includegraphics[width=\textwidth]{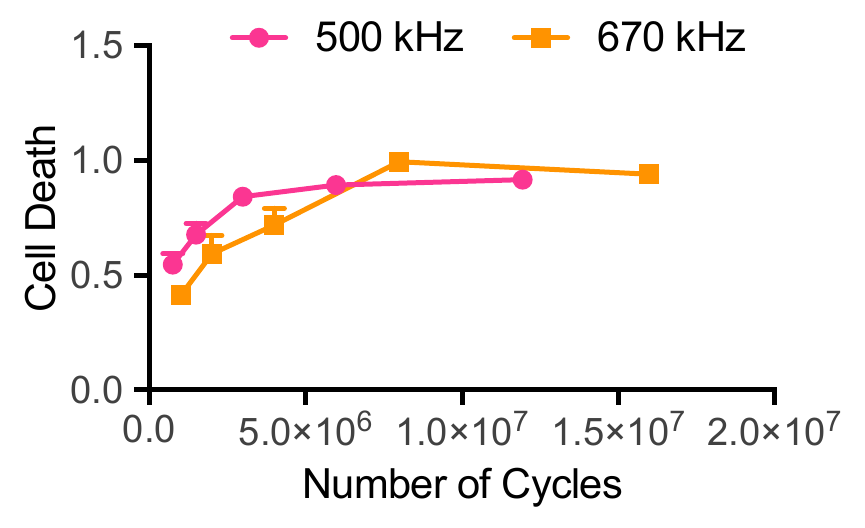}
		\caption{}
		\label{fig:celldeathvstimek562}
	\end{subfigure}
	\begin{subfigure}[b]{0.35\textwidth}
		\includegraphics[width=\textwidth]{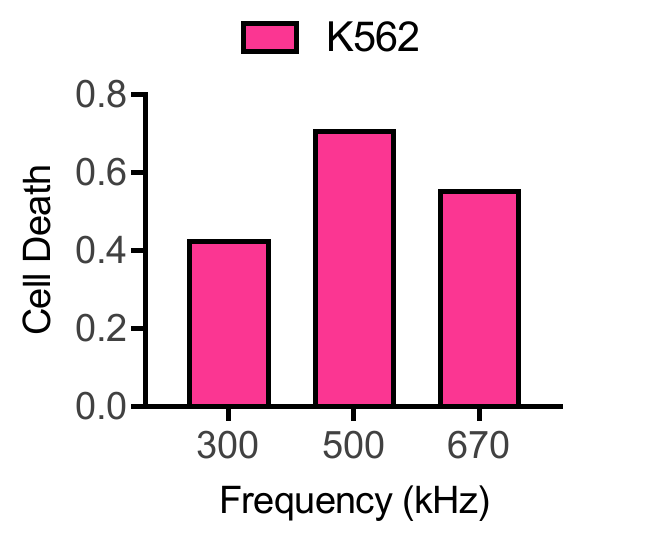}
		\caption{}
		\label{fig:celldeathvtimeu937}
	\end{subfigure}
    \caption{Tests of cancerous cell K562 at a free field pressure of 0.7 MPa, pulse duration of 100 ms and duty cycle of 10\%, showing the effect of frequency and number of cycles. (A) Cell death vs number of cycles (B) Cell death at 1.8 million cycles. Unpublished data from Mittelstein {\sl et al.} \cite{Mittelstein:2019}. }\label{1HYBrW}
\end{figure}

\subsection{Cell-death data}

The experimental study of Mittelstein {\sl et al.} \cite{Mittelstein:2019} reveals that LIPUS conditions at specific frequencies and pulsing parameters can indeed achieve cell-selective cytodisruption. This capability to tune ultrasound parameters to cause selective disruption in cancer cells while sparing healthy cells appears to be a novel finding and fits with many of the predictions of the oncotripsy theory. The morphology, type and related disease for each cell line are listed in Table~\ref{tbl:varprop}.

Figure~\ref{v0EZTl} demonstrates that cells can have varying responses to ultrasound depending on the ultrasound waveform.  All data points in this figure represent cell death assessed using LIVE/DEAD assays after exposure to an equal dosage of acoustic energy, though administered with different signal frequencies and pulse durations.  These tests were all performed on cells in suspension for an exposure time of 60 seconds, a duty cycle of 10\%, and in a spatial standing wave setup with a free field pressure of 0.7 MPa. Remarkably, high cell-death rates are observed for both the cancerous K-562 and U937 lines at 500 kHz signal frequency and 20 ms pulse duration while, under identical conditions, the control T-cells remain nearly unaffected (see Fig.~\ref{v0EZTl}b). These observations bear out the oncotripsy effect, as a frequency-dependent resonant response---and eventual death---of cells under harmonic excitation, and its selectivity.

The data in Fig.~\ref{v0EZTl} also shows a strong dependence of the cell response on pulse duration, with cell death enhanced at higher pulse durations. We take this dependence to suggest that the cell response is the result of two competing effects with vastly different characteristic times: damage accumulation during the on-part of the cycle and cell repair and healing during the entire time of exposure. The efficiency of the duty cycle may then be expected to depend sensitively on the relative values of the pulsing period and the characteristic times for damage accumulation and healing.

Figure~\ref{1HYBrW} shows data from tests of cancerous K562,
%that we will model as being exposed to acoustic pressure waves with a peak negative pressure of 0.7 MPa in free water and a pulse duration of 100 ms,
showing the effect of frequency and the number of cycles. In all cases, the pulse duration is 10\% of the total pulse repetition period, or a duty factor of 0.1.  As may be seen from these figures, cell death does not occur instantly but requires a certain exposure time to occur. We take this observation to suggest that death occurs by a process of damage accumulation over many insonation cycles. It is also evident from the figures that some cells die relatively early whereas others require considerably large number of cycles to die. These observations are suggestive of a broad variability in the susceptibility of the cell population to LIPUS.

\section{Oncotripsy model}

We proceed to develop a theoretical framework in which to understand and rationalize the preceding observations. The framework explored in this work is based on the following assumptions:
\begin{itemize}
\item[i)] For cells in suspension subjected to ultrasound, the aqueous medium damps out and suppresses the outer membrane vibrations, which translates rigidly at the particle velocity of the water.
\item[ii)] The internal structures of the cell, including its nucleus, respond as a resonator and vibrate in sync with the applied ultrasound.
\item[iii)] For sufficiently large pulse amplitudes, the cytoskeleton sustains cumulative mechanical damage that increases with successive cycles.
\item[iv)] At all times during exposure to ultrasound, the cytoskeleton can repair itself at a rate proportional to the level of damage sustained.
\item[v)] The cell ceases to be viable and dies when the amount of cumulative damage to the cytoskeleton exceeds a critical threshold.
\end{itemize}
The various elements of the theory are next developed in turn.

\begin{figure*}
    \centering
    \includegraphics[width=0.8\textwidth]{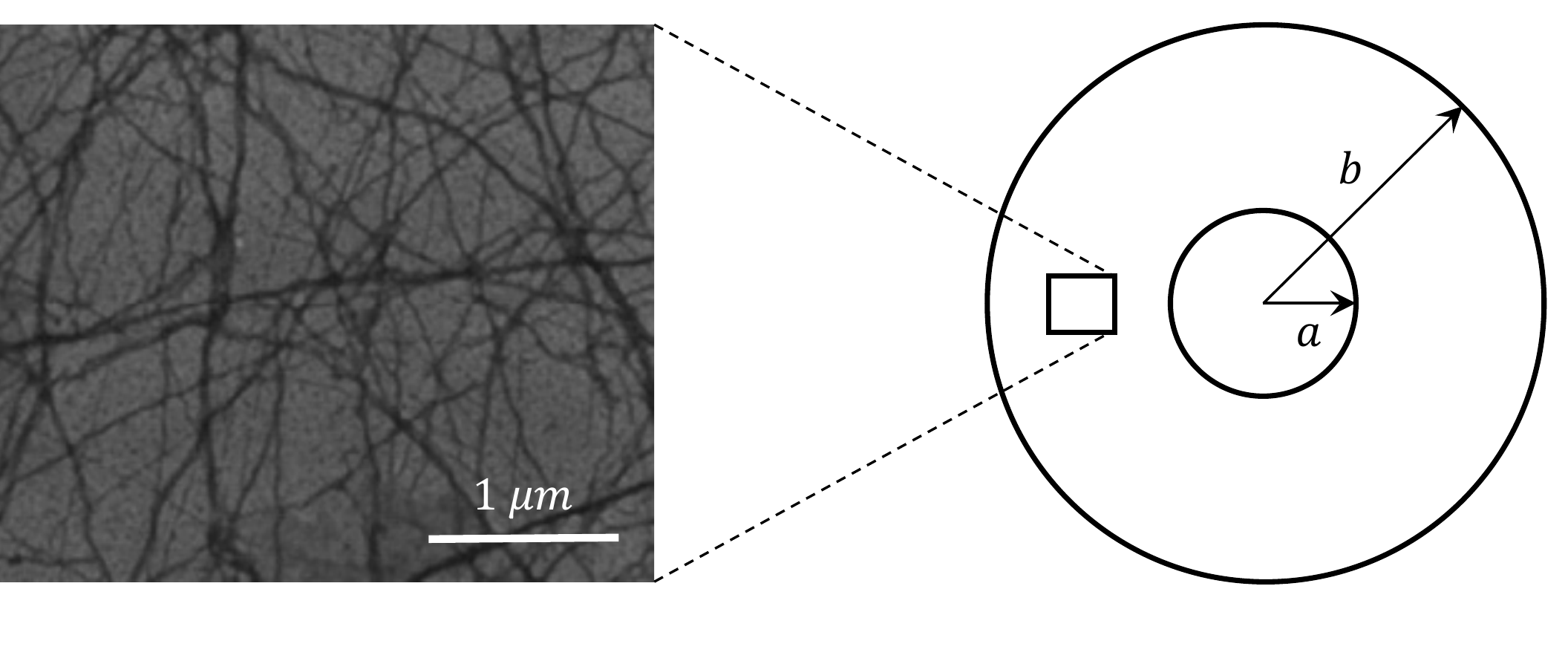}
    \caption{\small Schematic of the computational model, consisting of a random three-dimensional network of filaments spanning a rigid and heavy nucleus and a cell membrane oscillating rigidly with the surrounding fluid. Left: Electron micrograph of F-actin (adapted from \cite{Gardel:2008}).} \label{fD1aZT}
\end{figure*}

\subsection{Three-dimensional structure}

In mammalian cells, the nucleus, as the largest cellular organelle, occupies about $10$\,\% of the total cell volume \cite{Alberts:2002, Lodish:2004}. It is surrounded by the cytosol, a viscoelastic solid containing several subcellular structures such as the Golgi apparatus, the mitochondrion, and the endoplasmic reticulum. The cytosol and other organelles contained within the plasma membrane, for instance mitochondria and plastids, form the so-called cytoplasm. The nucleus is bounded by the nuclear envelope and contains the nucleoplasm, a viscoelastic solid similar in composition to the cytosol. It furthermore comprises the nucleolus, which constitutes the largest structure within the nucleus and consists of proteins and RNA. In the present work, we neglect the organelles within the cytosol, which is idealized as a uniform viscous matrix containing the cytoskeleton. The nucleus is likewise idealized as rigid and we omit explicit consideration of the nucleoplasm. Given the focus on cytoskeletal dynamics, we additionally neglect the effect of the nuclear and cellular membranes.

\subsection{Cytoskeleton elasticity}

The cytoskeleton is a system of filaments in the cell that radiates from the nucleus and is anchored at the plasma membrane. In eukaryotic cells, the filament network has three major components: microtubules, intermediate filaments and microfilaments, Fig.~\ref{fD1aZT}a. Microfilaments are polymers of the protein actin, microtubules are composed of the protein tubulin and intermediate filaments are composed of various proteins, depending on the type of cell. The cytoskeleton confers elasticity to the cell, mediates the movement of the cells, helps to support the cytoplasm and responds against external mechanical stimuli.  In particular, microfilaments and intermediate filaments act as cables to support tension loads while microtubules act as beams in compression \cite{Li2011}, in analogy to tensegrity structures \cite{Ingber2003, Kardas2013, Barreto2013, Li2005}.

According to the {\sl network theory of elasticity} in statistical mechanics \cite{Weiner:1983, Flory:1989}, the cytoskeleton may be modeled as an amorphous network of cross-linked fibers. The fibers consist of many freely-jointed segments and are far from full extension. It is further assumed that the cross-linking points move according to the local macroscopic deformation. In addition, the cytoskeleton is assumed to be embedded in a viscous matrix. A standard analysis (cf., e.~g., \cite{Weiner:1983}) then gives the free-energy density per unit volume of the network as
\begin{equation}\label{FIWY7N}
    A({F}, T)
    =
    \frac{\mu(T)}{2} K_{IJ} (C_{IJ} + C^{-1}_{IJ}) ,
\end{equation}
up to inconsequential additive constants. In (\ref{FIWY7N}), $\mu(T)$ is a temperature-dependent shear modulus, ${F}$ is the local deformation gradient, ${C} = {F}^T{F}$ is the right Cauchy-Green deformation tensor and $T$ is the absolute temperature (cf., e.~g., \cite{Weiner:1983, Marsden:1994} for background on continuum mechanics). An analysis of the configurational entropy of the fibers \cite{Weiner:1983, Flory:1989} gives the shear modulus as
\begin{equation}
    \mu(T) = \frac{2 n l^2}{b^2} k_B T ,
\end{equation}
where $n$ is the number of fibers per unit volume, $b$ is the segment length, $l$ is the end-to-end distance of the fibers and $k_B$ is Bolzmann's constant. In addition, the structure tensor ${K}$ in (\ref{FIWY7N}) follows as
\begin{equation}\label{gTjtN9}
    K_{IJ}
    =
    \int_{S^2}
        p({\xi}) \xi_I \xi_J
    \, d\Omega ,
\end{equation}
where $\xi$ is the unit vector pointing from one end of the fiber to the other, or {\sl fiber direction}, $p({\xi})$ is the fraction of chains in the ensemble of direction $\xi$, $S^2$ is the unit sphere and $d\Omega$ is the element of solid angle. The density $p({\xi})$ is subject to the normalization condition
\begin{equation}
    \int_{S^2}
    p({\xi})
    \, d\Omega
    =
    1 .
\end{equation}
The distribution function $p({\xi})$ describes the structure of the cytoskeletal network and is assumed fixed and known. For instance, Smolyakov {\sl et al.} \cite{Smolyakov:2016} used single-cell force spectroscopy to test mechanical properties of four breast cancer 11 cell lines and found that the most invasive cells, MDA-MB231, contain actin fibers that are distributed randomly throughout the cell without any particular structure of preferred direction. For an {\sl isotropic} fiber distribution of this type, $p = \nicefrac{1}{4\pi}$, and the structure tensor (\ref{gTjtN9}) reduces to the identity. Under these conditions, the free-energy density (\ref{FIWY7N}) specializes to
\begin{equation}
    A(F, T) = {{\mu(T)}\over{2}} \big( {\rm tr}({C}) +\emph{} {\rm tr}({C}^{-1}) \big) ,
\end{equation}
where ${\rm tr}$ denotes the matrix trace.

\subsection{Cytoskeletal damage and healing}

The experimental observations of Mittelstein {\sl et al.} \cite{Mittelstein:2019} for cells in suspension, Section~\ref{i9UQnX}, reveal that cell death requires the application of a large number (millions) of insonation pulses, which in turn suggests that, under the conditions of the experiment, cell death is the result of a process of slow damage accumulation. Indeed, Mizrahi {\sl et al.} \cite{Mizrahi:2012} observed that, whereas the cytoskeletal actin fibers are catastrophically disrupted under the action of ultrasound stimulation of sufficiently high intensity, Fig.~\ref{gGZ1YS}, under low-intensity ultrasound cellular responses exhibit gradual damage accumulation and sometimes complete recovery following insonation cessation. Confocal microscopy of CT-26 cells assessed after LIPUS treatment reported in \cite{Mittelstein:2019} also reveals that LIPUS cytodisruption is coupled with persistent cytoskeletal disruption, cf.~Fig.~\ref{x6sPlV}.

Whereas cytoskeletal elasticity has been extensively studied in the past, processes of damage accumulation in the cytoskeleton under LIPUS actuation, or high-cycle cell fatigue, appear to be as yet poorly understood. Building on past work on failure of polymer networks \cite{Balzani:2012, Heyden:2015a, Heyden:2015b}, we develop a model of cumulative cell damage that accounts for the gradual disruption and repair of cytoskeletal fibers. This competition between disruption (`death') and repair (`birth') is a classical example of a `birth-death' process in evolutionary dynamics, cf., e.g., \cite{Nowak:2006}.

We assume that the mechanism of damage accumulation to the cytoskeleton is the progressive disruption of the actin fibers. In order to account for the attendant loss of stiffness, we introduce a damage variable $q(\xi)$ ranging from $0$ to $1$ such that $q(\xi) = 0$ when all the fibers with direction $\xi$ are intact and $q(\xi) = 1$ when all the fibers with direction $\xi$ are broken. We additionally assume that the breaking of the fibers requires a certain energy to be supplied. We represent these effects by means of a free-energy density of the form
\begin{equation}\label{Bo43Ae}
    A({F}, T, q)
    =
    \int_{S^2}
        p({\xi})
        \Big(
        \frac{\mu(T)}{2}
        (1 - q(\xi))^2
        \big(\lambda^2(\xi) + \lambda^{-2}(\xi) - 2\big)
        +
        \frac{\beta}{2} q^2(\xi)
        \Big)
    \, d\Omega ,
\end{equation}
where
\begin{equation}
    \lambda(\xi)
    =
    \sqrt{C_{IJ} \xi_I \xi_J}
\end{equation}
is the stretch ratio of the fibers of direction $\xi$ and $\beta$ is a constant. We note from (\ref{Bo43Ae}) that the effect of a damage field $q(\xi)$ is to decrease the free-energy density of the fibers of direction $\xi$ by a factor $(1 - q(\xi))^2$ at an energy cost of $(\beta/2) q^2(\xi)$. Additionally, damage relaxes the stresses in the network by reducing the stiffness of the fibers. Evidently, in the absence of damage, $q(\xi) = 0$, (\ref{Bo43Ae}) reduces to (\ref{FIWY7N}), as required.

Following the method of Coleman and Noll \cite{Coleman:1963}, the thermodynamic driving forces for damage follow as
\begin{equation}\label{VAk6da}
    f(\xi)
    =
    -
    \frac{\partial A}{\partial q(\xi)}
    =
    p({\xi})
    \Big(
        \mu(T)
        (1 - q(\xi))
        \big(\lambda^2(\xi) + \lambda^{-2}(\xi) - 2\big)
        -
        \beta q(\xi)
    \Big) .
\end{equation}
We see from this expression that, by the choice (\ref{Bo43Ae}) of free-energy density, the driving force (\ref{VAk6da}) comprises two terms. The first term represents the energy-release rate due to the disruption of the fibers and, therefore, promotes damage. The second term represents the energetic cost of disrupting the fibers, which hinders damage and promotes healing. Assuming linear kinetics, we obtain the damage evolution law
\begin{equation}\label{dn28B7}
    \alpha \,
    \dot{q}(\xi)
    =
    f(\xi) ,
\end{equation}
where $\alpha$ is a kinetic coefficient.

The kinetic relation (\ref{dn28B7}), in combination with the driving forces (\ref{VAk6da}), define an evolution of the cytoskeletal state as a balance between 'birth' and 'death' processes. Thus, the energy-release term $\mu(T)(1 - q(\xi))\big(\lambda^2(\xi) + \lambda^{-2}(\xi) - 2\big)$ in the driving force induces progressive damage ('death') of the fiber population proportionally to the energy $\mu(T)\big(\lambda^2(\xi) + \lambda^{-2}(\xi) - 2\big)$ of the fibers. The additional factor $(1 - q(\xi))$ brings the driving force to zero at full damage $q(\xi) = 1$ and ensures that $q(\xi) \leq 1$ at all times. By contrast, the energetic cost term $- \beta q(\xi)$ in the driving force tends to restore ('birth') the fiber population and thus accounts for healing. Built into the form of (\ref{VAk6da}) is the assumption that the rate of healing is proportional to the extent of damage. In particular, the healing rate vanishes for $q(\xi) = 0$, which it ensures that $q(\xi) \geq 0$ at all times.

\subsection{Cell viscosity}

Another source of resistance to cell deformation arises from the viscosity of the cytoplasm. This viscosity damps resonant vibrations within the cell and limits their amplitude. On average, the cytoplasm viscosity does not differ significantly from that of water \cite{Fushimi:1991, LubyPhelps:1993}, but the distribution of intracellular viscosity is highly heterogeneous. Full maps of subcellular viscosity have been successfully constructed via fluorescent ratiometric detection and fluorescence lifetime imaging \cite{Liu:2014}. However, this degree of detail is beyond the scope of this study. Instead, we assume an average viscosity uniformly distributed over the cytoplasm. Further assuming linear viscosity, the viscous Cauchy stress in the cytoplasm follows as
\begin{equation}\label{R8j8SA}
    \sigma_{ij}
    =
    \eta
    ( v_{i,j} + v_{j,i} )
    +
    \left( \kappa - \frac{2}{3} \eta \right)
    {\rm div} \, v \, \delta_{ij} ,
\end{equation}
where $\eta$ is the shear viscosity, $\kappa$ is the bulk viscosity, $v$ is the velocity field, a comma denotes partial differentiation and ${\rm div} \, v$ is the divergence of the velocity field.

\subsection{Reduced model}

The preceding model of cytoplasm elasticity, damage, healing and viscosity can be taken as a basis for a fully three-dimensional analysis of cell motion, e.~g., by means of the finite-element method, cf.~\cite{EFS2019}. However, parametric and sensitivity studies are greatly facilitated by reduced models. We develop a reduced dynamical model of cell deformation and damage based on the following assumptions:
\begin{itemize}
\item[i)] Spherical geometry of cell and nucleus.
\item[ii)] Rigid translational motion of the cell membrane.
\item[iii)] Heavy and rigid nucleus.
\item[iv)] {\sl Ans\"atze} for the cytoplasm deformation and damage fields.
\end{itemize}
We note that, under the conditions of interest here, a Rayleigh treatment of the acoustic scattering problem is justified in view of the large wavelength of the ultrasound waves compared to the cell size.

\begin{figure*}
	\centering
	\begin{subfigure}[b]{0.45\textwidth}
		\centering
		\includegraphics[totalheight=0.3\textheight]{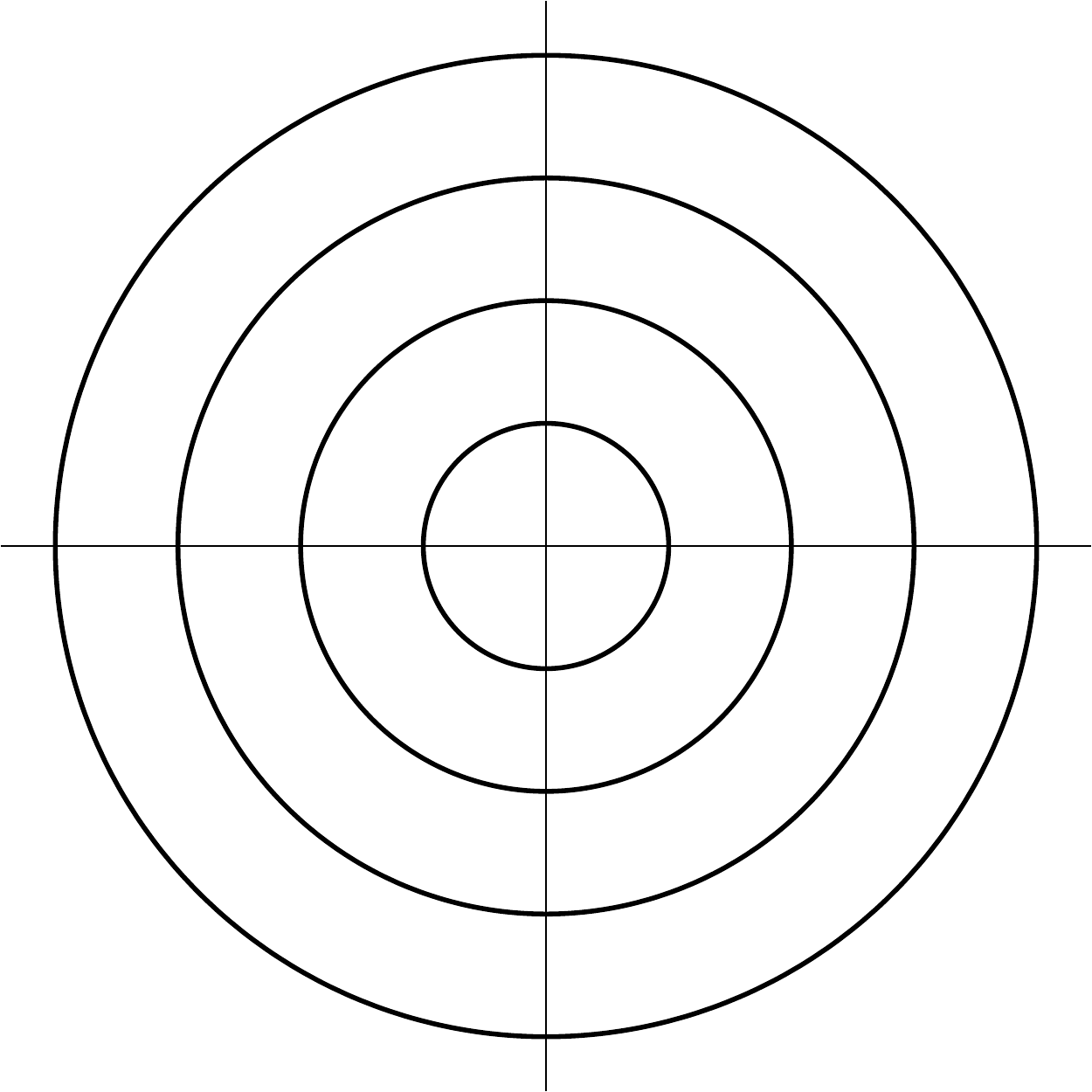}
		\caption[]{}
	\end{subfigure}
	%\quad
	\begin{subfigure}[b]{0.45\textwidth}
		\centering
		\includegraphics[totalheight=0.3\textheight]{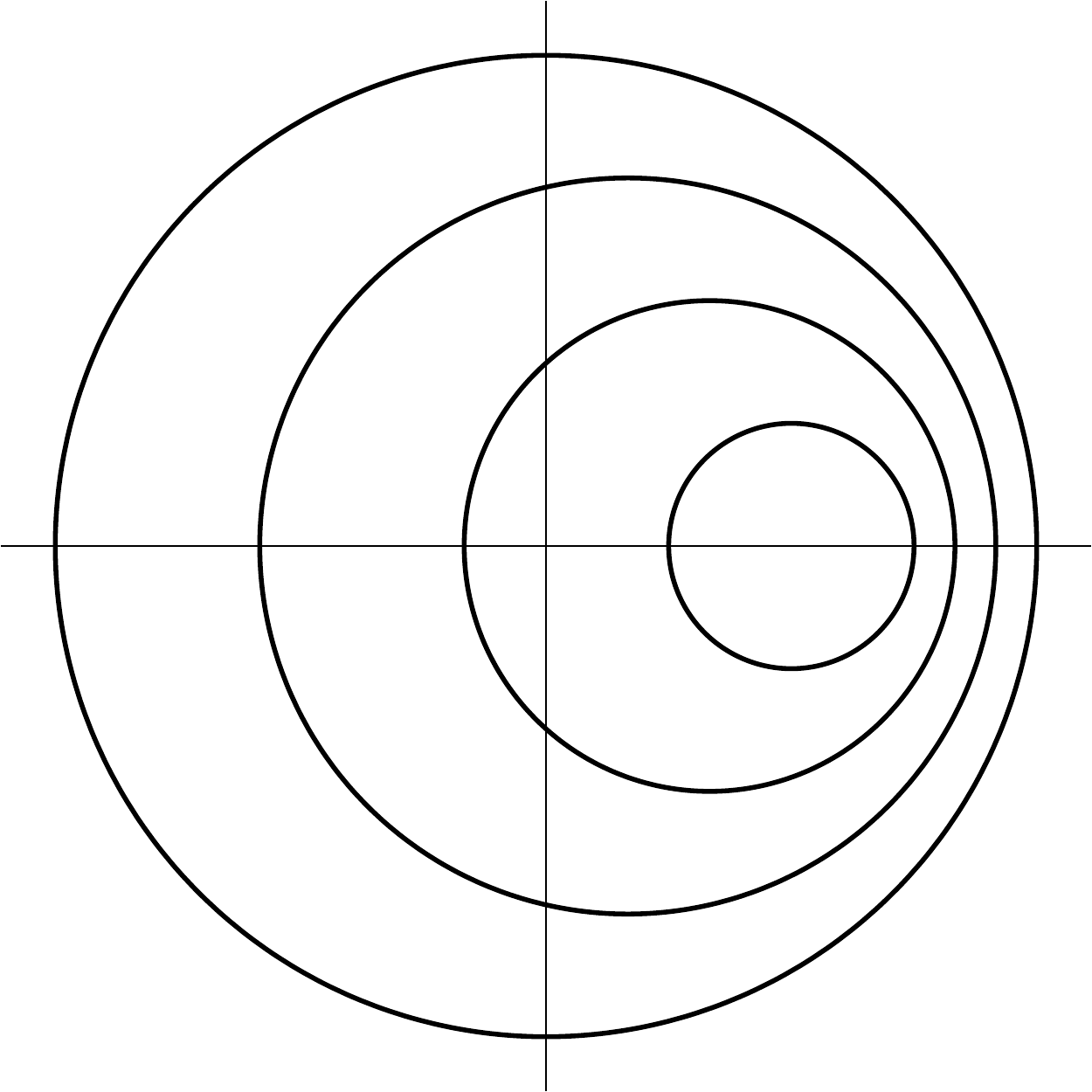}
		\caption[]{}
	\end{subfigure}
	\caption{\small Deformation {\sl ansatz} used in model reduction. a) Cross section of the reference configuration of the cell, showing nucleus (inner circle) and two concentric material spheres to aid in the visualization of the deformation. b) Deformed configuration of the cell after a displacement of the nucleus.} \label{wrUt5a}
\end{figure*}

We specifically consider a spherical cell of radius $b$ containing a concentric spherical nucleus of radius $a$. We assume that the cell moves under the action of planar waves and executes a translational motion according to the particle velocity of the aqueous medium. We attach a moving cartesian reference frame to the center of the cell such that the $x_3$ axis is aligned with the direction of motion. We additionally introduce a spherical coordinate system $(r,\varphi,\theta)$, such that
\begin{equation}
    x_1 = r \, \sin\theta \, \cos\varphi ,
    \qquad
    x_2 = r \, \sin\theta \, \sin\varphi ,
    \qquad
    x_3 = r \, \cos\theta ,
\end{equation}
where $r$ is the radius, $\varphi$ is the azimuthal angle and $\theta$ the inclination. In these spherical coordinates, the domain of the cytoplasm in its undeformed configuration is $\varphi \in [0,2\pi)$, $\theta \in [0,\pi)$ and $r \in [a,b]$. The nucleus is assumed to translate rigidly through a time-dependent displacement $u(t)$ relative to the cell membrane. In addition, a material point in the cytoplasm initially at location $(x_1,x_2,x_3)$ in the undeformed configuration is assumed to be at location
\begin{equation}\label{D2ImDc}
    y_1 = x_1 ,
    \qquad
    y_2 = x_2 ,
    \qquad
    y_3 = x_3 + \frac{b-r}{b-a} \, u(t) ,
\end{equation}
following the displacement of the nucleus. In this {\sl ansatz}, a spherical material shell of radius $r$ in the undeformed configuration translates rigidly to another spherical shell of the same radius centered at $u(t) \, (b-r)/(b-a)$ following the displacement of the nucleus, cf.~Fig.~\ref{wrUt5a}.

\subsubsection{Dynamics without damage}
Inserting this {\sl ansatz} into the free-energy density (\ref{FIWY7N}) and assuming small relative displacements $u(t)$, we obtain, after a trite calculation,
\begin{equation}
    A
    =
    \frac{\mu}{2}
    \frac{u^2(t)}{(b-a)^2}
    (3 + \cos(2\theta)) ,
\end{equation}
and the total free energy of the cytoskeleton evaluates to
\begin{equation}\label{8Re3Dz}
    \mathcal{A}(u(t))
    =
    \int_0^{2\pi}\int_0^\pi\int_a^b
        A r^2 \sin\theta
    \, dr \, d\theta \, d\varphi
    =
    \frac{16\pi}{9} (b^3 - a^3) \mu \frac{u^2(t)}{(b-a)^2} ,
\end{equation}
which, in the absence of damage, supplies a potential for the relative displacement of the nucleus. Likewise, the velocity field of the cytoplasm follows by time differentiation of the {\sl ansatz} (\ref{D2ImDc}), with the result
\begin{equation}
    v_1 = 0 ,
    \qquad
    v_2 = 0 ,
    \qquad
    v_3 = \frac{b-r}{b-a} \, \dot{u}(t) .
\end{equation}
Inserting this velocity field into the viscosity law (\ref{R8j8SA}) and assuming small relative displacements of the nucleus gives, after a straightforward calculation, the dissipation per unit undeformed volume
\begin{equation}
    D
    =
    \frac{1}{2} \sigma_{ij} v_{i,j}
    =
    \frac{1}{24}
    \big(5 \eta + 6 \kappa - (\eta - 6 \kappa) \cos(2\theta)\big)
    \frac{\dot{u}^2(t)}{(b-a)^2} ,
\end{equation}
and the total dissipation follows as
\begin{equation}\label{8SkLQD}
    \mathcal{D}(\dot{u}(t))
    =
    \int_0^{2\pi}\int_0^\pi\int_a^b
        D r^2 \sin\theta
    \, dr \, d\theta \, d\varphi
    =
    \frac{2\pi}{27}
    (b^3 - a^3)
    (4 \eta + 3 \kappa)
    \frac{\dot{u}^2(t)}{(b-a)^2} .
\end{equation}
Finally, the total kinetic energy of the cell follows as
\begin{equation}\label{r3DrLT}
    \mathcal{K}(t,\dot{u}(t))
    =
    \frac{1}{2}
    \Big(
        m_0
        +
        \frac{2\pi}{15} \rho (b-a)(6 a^2 + 3 a b + b^2)
    \Big)
    (v(t) + \dot{u}(t))^2 ,
\end{equation}
where $m_0$ is the mass of the nucleus, $\rho$ is the density of the cytoplasm and $v(t)$ is the prescribed velocity of the cell membrane. An appeal to the Lagrange-D'Alembert principle gives the equation of motion
\begin{equation}\label{piT2pi}
    \frac{d}{dt} \frac{\partial \mathcal{K}}{\partial \dot{u}}(t,\dot{u}(t))
    +
    \frac{\partial \mathcal{D}}{\partial \dot{u}}(\dot{u}(t))
    +
    \frac{\partial \mathcal{A}}{\partial u}(u(t))
    =
    0 .
\end{equation}
Inserting (\ref{8Re3Dz}), (\ref{8SkLQD}) and (\ref{r3DrLT}) into (\ref{piT2pi}), we obtain
\begin{equation}\label{v5tHov}
    m \ddot{u}(t)
    +
    c \dot{u}(t)
    +
    k u(t)
    =
    -
    m \dot{v}(t) ,
\end{equation}
where
\begin{equation}
    m
    =
    m_0
    +
    \frac{2\pi}{15} \rho (b-a)(6 a^2 + 3 a b + b^2) ,
    \quad
    c
    =
    \frac{4\pi}{27}
    \frac{b^3 - a^3}{(b-a)^2}
    (4 \eta + 3 \kappa) ,
    \quad
    k
    =
    \frac{32\pi}{9} \frac{b^3 - a^3}{(b-a)^2} \mu ,
\end{equation}
are the total mass, damping coefficient and stiffness of the cell. Eq.~(\ref{v5tHov}) describes a damped and forced harmonic oscillator, with the material velocity $v(t)$ of the aqueous medium supplying the forcing.

\subsubsection{Dynamics with damage}
Suppose now that the cell undergoes damage. In general, damage patterns may be expected to arise at two levels: inhomogeneously over the cytoplasm; and damage along preferential fiber directions at every material point. Such degree of complexity requires a full three-dimensional analysis for its elucidation, cf.~\cite{EFS2019}. In order to simplify the dynamics, we simply assume that damage is isotropic at all material points, i.~e., the damage parameter $q$ is independent of direction $\xi$; and independent of position over the cytoskeleton. By this simple {\sl ansatz}, the state of damage of the cell is characterized by a single state variable $q(t)$. An immediate extension of (\ref{8Re3Dz}) then gives the total free energy of the cell as
\begin{equation}\label{Ga1hop}
    \mathcal{A}(u(t), q(t))
    =
    \frac{16\pi}{9} (b^3 - a^3) (1-q(t))^2 \mu \frac{u^2(t)}{(b-a)^2}
    +
    \frac{4\pi}{3} (b^3 - a^3) \frac{\beta}{2} q^2(t) .
\end{equation}
Likewise, the total dissipation (\ref{8SkLQD}) extends to
\begin{equation}\label{DowI5R}
    \mathcal{D}(\dot{u}(t), \dot{q}(t))
    =
    \frac{2\pi}{27}
    (b^3 - a^3)
    (4 \eta + 3 \kappa)
    \frac{\dot{u}^2(t)}{(b-a)^2}
    +
    \frac{4\pi}{3} (b^3 - a^3) \frac{\alpha}{2} \dot{q}^2(t) .
\end{equation}
The Lagrange-D'Alembert principle then gives the coupled equations
\begin{subequations}\label{5Wt8Ag}
\begin{align}
    &
    \frac{d}{dt} \frac{\partial \mathcal{K}}{\partial \dot{u}}(t,\dot{u}(t))
    +
    \frac{\partial \mathcal{D}}{\partial \dot{u}}(\dot{u}(t), \dot{q}(t))
    +
    \frac{\partial \mathcal{A}}{\partial u}(u(t), q(t))
    =
    0 ,
    \\ &
    \frac{\partial \mathcal{D}}{\partial \dot{q}}(\dot{u}(t), \dot{q}(t))
    +
    \frac{\partial \mathcal{A}}{\partial q}(u(t), q(t))
    =
    0 .
\end{align}
\end{subequations}
Inserting (\ref{Ga1hop}), (\ref{DowI5R}) and (\ref{r3DrLT}) into (\ref{5Wt8Ag}), we now obtain
\begin{subequations}\label{frEpi9}
\begin{align}
    &   \label{kIC2St}
    m \ddot{u}(t)
    +
    c \dot{u}(t)
    +
    (1 - q(t))^2 k u(t)
    =
    -
    m \dot{v}(t) ,
    \\ &    \label{vot6Lx}
    {n} \dot{q}(t) + {d} q(t) = (1-q(t)) % \frac{k}{2}
    k u^2(t) ,
\end{align}
\end{subequations}
with $m$, $c$ and $k$ as before and
\begin{equation}
    {n} = \frac{4\pi}{3}(b^3-a^3) \, \alpha,
    \qquad
    {d} = \frac{4\pi}{3}(b^3-a^3) \, \beta.
\end{equation}
The first of these equations represents a forced and damped harmonic oscillator in which the stiffness depends on the instantaneous state of damage. The second governs the kinetic evolution of the damage state, including damage accumulation and healing.

\begin{figure}
	\centering
	\begin{subfigure}[]{0.45\textwidth}
		\centering
		\includegraphics[width = 0.99\textwidth]{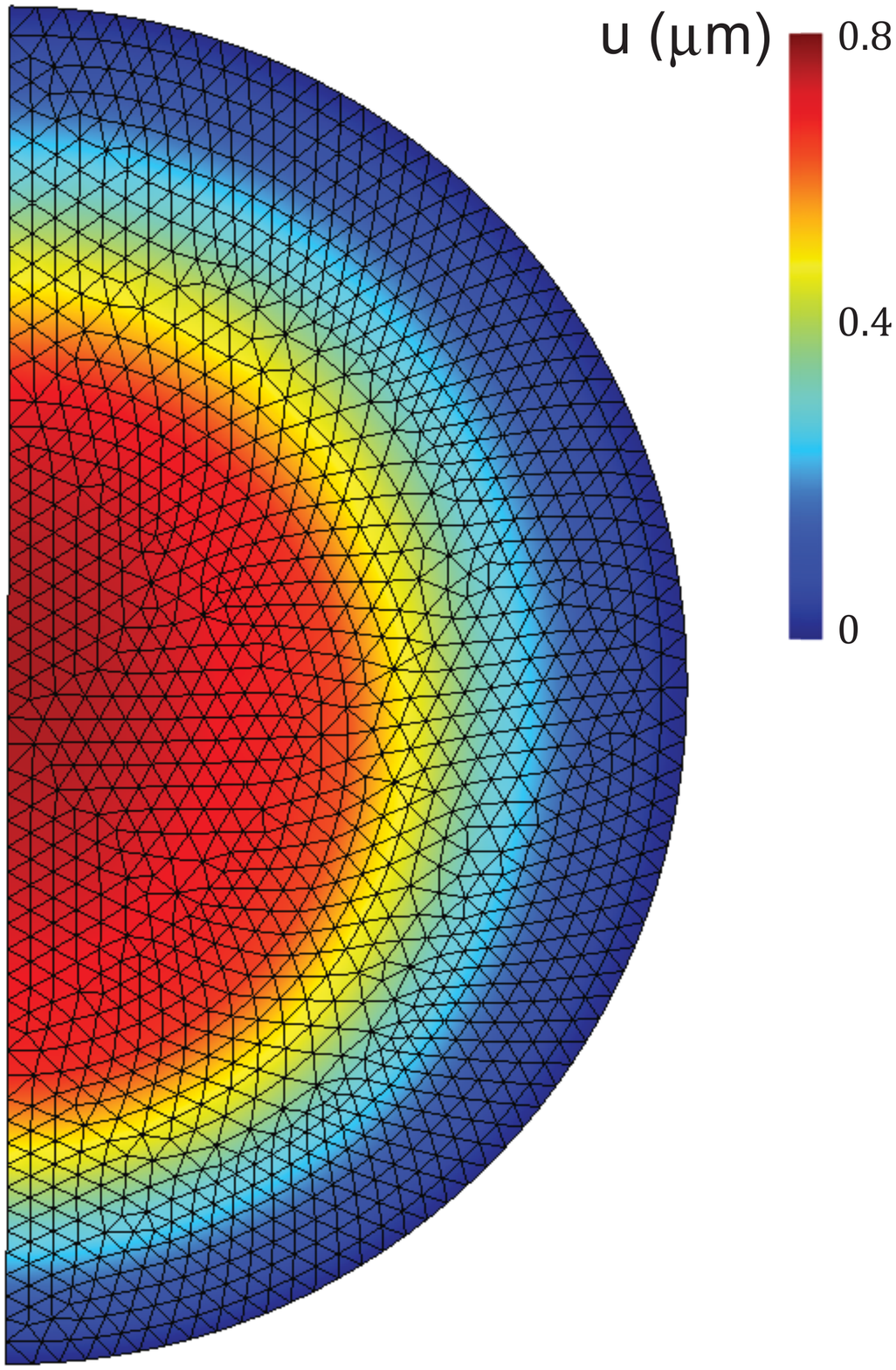}
		\caption[]{}
	\end{subfigure}
	\begin{subfigure}[]{0.45\textwidth}
		\centering
		\includegraphics[width = 0.99\textwidth]{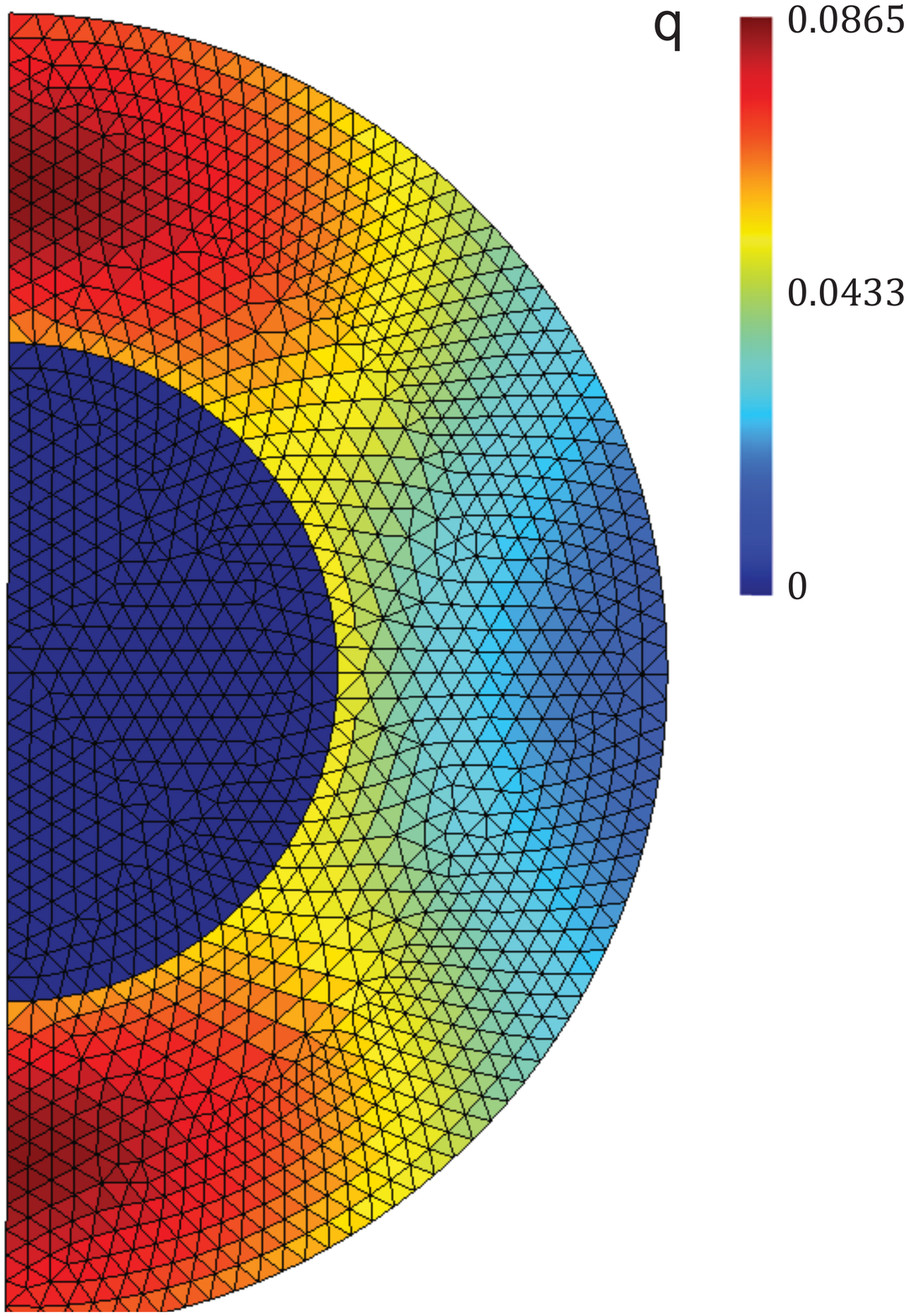}
		\caption[]{}
	\end{subfigure}
    \caption{\small Axisymmetric finite-element results at $0.1$ ms exposure using the full three-dimensional damage model \cite{EFS2019}. The cell is insonated at 1.4MPa focal pressure and 500 kHz frequency. a) Magnitude of axial displacement in microns. b) Damage averaged over all fiber directions.
} \label{Ze3esi}
\end{figure}

The accuracy of the reduced model just derived can be assessed by means of comparisons with finite-element implementations of the full model. Fig.~\ref{Ze3esi} shows a typical axisymmetric calculation in which a cell is insonated at 1.4MPa focal pressure and 500 kHz frequency over $0.1$ ms \cite{EFS2019}. As may be seen from the figure, the damage to the cytoskeleton is localized at poles of the cell. Despite this patterning, the nuclear displacements and average cytoskeletal damage predicted by the reduced model are found to be within 7\% of the full-field finite-element calculations. Given the level of observational error, this accuracy may reasonably be deemed adequate for all practical purposes. Further details of the error analysis may be found in \cite{EFS2019}.

\subsection{WKB dynamics}

Under the conditions of interest here, the dynamics described by system (\ref{5Wt8Ag}) is characterized by two disparate time scales: the period of oscillation and the characteristic time for damage evolution, the former much smaller than the latter. This two-time structure suggests analyzing the problem by means of WKB asymptotics \cite{Bender:1978}.

We consider a generic duty cycle such as shown inset in Fig.~\ref{8w5RPa}a, starting at time $t_0$ and consisting of an on-period ending at time $t_1$ and an off-period ending at time $t_2$. The duration of the on-period, or pulse duration, is $T_1=t_1-t_0$, the duration of the off cycle, or listening time, is $T_2=t_2-t_1$ and the total duration of the duty cycle, or pulse repetition period, is $T=t_2-t_0$. We specifically assume harmonic excitation of the form
\begin{equation} \label{7refRO}
    v(t) = V {\rm e}^{i \omega t}  ,
\end{equation}
during the on-period and $v(t) = 0$ during the off-period. In (\ref{7refRO}), $V$ is a complex amplitude and $\omega$ is the insonation frequency.

We begin by analyzing the equation of motion (\ref{kIC2St}), which we rewrite in the form
\begin{equation}
    \ddot{u}(t)
    +
    2 \zeta \omega_0 \dot{u}(t)
    +
    (1 - q(t))^2 \omega_0^2 u(t)
    =
    -
    \dot{v}(t) ,
\end{equation}
where $\omega_0 = \sqrt{k/m}$ is the {\sl natural frequency} of the undamaged cell and $\zeta$ is the {\sl damping ratio}. During the on-period of the duty cycle, we have
\begin{equation}\label{3AmoPu}
    \ddot{u}(t)
    +
    2 \zeta \omega_0 \dot{u}(t)
    +
    (1 - q(t))^2 \omega_0^2 u(t)
    =
    -
    i \omega V {\rm e}^{i \omega t} ,
\end{equation}
where, for convenience, we extend the equation to the complex domain. Assume now that the period of oscillation $T_0 = 2\pi/\omega_0$ is much smaller than the pulse duration $T_1$. Assume additionally that the frequency $\omega$ of insonation is comparable to $\omega_0$. Finally, suppose that the variation of the damage state variable $q(t)$ is slow and on the scale of the pulse duration $T_1$. Under these conditions, the solution $u(t)$ can be obtained by performing a WKB asymptotic analysis in the small parameter $T_0/T_1$. We note that, for fixed $q(t)$, eq.~(\ref{3AmoPu}) is a linear second-order ordinary differential equation and, therefore, its solution is the sum of the general homogeneous solution and a particular solution. Owing to the presence of damping, with damping coefficient $\zeta$ of $O(1)$, the homogeneous solution decays on the scale of $T_0$ and can be safely neglected. We seek a particular equation of the form
\begin{subequations}
\begin{align}
    &    \label{beQl4o}
    u(t) = A(t) {\rm e}^{i \omega t} ,
    \\ &
    \dot{u}(t)
    =
    ( \dot{A}(t) + i \omega A(t) )
    {\rm e}^{i \omega t} ,
    \\ &
    \ddot{u}(t)
    =
    \big(
        \ddot{A}(t)
        +
        2 i \omega \dot{A}(t)
        -
        \omega^2 A(t)
    \big)
    {\rm e}^{i \omega t} .
\end{align}
\end{subequations}
Inserting these expressions into (\ref{3AmoPu}) and retaining leading-order terms only, we obtain
\begin{equation}\label{7tLYaD}
    -
    \omega^2 A(t)
    +
    2 i \zeta \omega_0 \omega A(t)
    +
    (1 - q(t))^2 \omega_0^2 A(t)
    =
    -
    i \omega V .
\end{equation}
Solving for the amplitude $A(t)$, we find
\begin{equation}
    A(t)
    =
    \frac
    {
        i \omega V
    }
    {
        \omega^2
        -
        (1 - q(t))^2 \omega_0^2
        -
        2 i \zeta \omega_0 \omega
    } .
\end{equation}
Finally, inserting into (\ref{beQl4o}) we obtain
\begin{equation}\label{4aPOFo}
    u(t)
    =
    \frac
    {
        i \omega V {\rm e}^{i \omega t}
    }
    {
        \omega^2
        -
        (1 - q(t))^2 \omega_0^2
        -
        2 i \zeta \omega_0 \omega
    } ,
\end{equation}
asymptotically as $T_0/T_1 \to 0$. We observe from (\ref{4aPOFo}) that the nucleus executes rapid oscillations relative to the cell membrane over the pulse duration in sync with the ultrasound excitation, with amplitude modulated by the damage variable $q(t)$.

Next, we turn to the damage evolution equation (\ref{vot6Lx}). Inserting solution (\ref{4aPOFo}) into (\ref{vot6Lx}) gives
\begin{equation}\label{drlB8l}
    {n} \dot{q}(t)
    +
    {d} q(t)
    =
    \frac
    {
        k (1-q(t)) \omega^2 |V|^2
    }
    {
        \big(
            \omega^2
            -
            (1 - q(t))^2 \omega_0^2
        \big)^2
        +
        4 \zeta^2 \omega_0^2 \omega^2
    } ,
\end{equation}
which is now fully expressed in terms of the damage variable $q(t)$. Conveniently, eq.~(\ref{drlB8l}) is separable and admits the explicit solution
\begin{equation}\label{KU8Ibr}
    t
    =
    t_0
    +
    \int_{q_0}^q
    \frac
    {
        {n} \, d\xi
    }
    {
        \dfrac
        {
            k (1-\xi) \omega^2 |V|^2
        }
        {
            \big(
                \omega^2
                -
                (1 - \xi)^2 \omega_0^2
            \big)^2
            +
            4 \zeta^2 \omega_0^2 \omega^2
        }
        -
        {d} \xi
    } ,
\end{equation}
where we write $q_0 = q(t_0)$. Alternatively, the equation of evolution (\ref{drlB8l}) can be recast in terms of dimensionless variables as
\begin{equation}\label{r0biGU}
    \frac{dq}{d\tau}(\tau)
    +
    q(\tau)
    =
    \frac
    {
        (1-q(t)) w^4 \varepsilon
    }
    {
        \big(
            w^2
            -
            (1 - q(\tau))^2
        \big)^2
        +
        4 \zeta^2 w^2
    } ,
\end{equation}
where
\begin{equation}
    \tau = \frac{t-t_0}{t_r},
    \qquad
    t_r = \frac{{n}}{{d}},
    \qquad
    w
    =
    \frac{\omega}{\omega_0},
    \qquad
    \varepsilon
    =
    \frac{k |V|^2}{{d} \omega_0^2}
    =
    \frac{m |V|^2}{{d} } ,
\end{equation}
whereupon (\ref{KU8Ibr}) becomes
\begin{equation}\label{tuS9Wr}
    \tau
    =
    \int_{q_0}^q
    \frac
    {
        d\xi
    }
    {
        \dfrac
        {
            (1-\xi) w^4 \varepsilon
        }
        {
            \big(
                w^2
                -
                (1 - \xi)^2
            \big)^2
            +
            4 \zeta^2 w^2
        }
        -
        \xi
    } .
\end{equation}
From this reparametrization, we observe that the evolution of damage depends on the following dimensionless parameters: i) The ratio of the elapsed time to the relaxation time $t_r$ for healing; ii) the ratio $w$ between the frequency of insonation and the undamaged natural frequency; iii) the energy deposited by insonation relative to the energy cost of repair; and iv) the cell damping ratio. It is also interesting to note that the damage state variable attains a steady-state maximum value $q_{\rm max}$ when
\begin{equation}
    q_{\rm max}
    =
    \frac
    {
        (1-q_{\rm max}) w^4 \varepsilon
    }
    {
        \big(
            w^2
            -
            (1 - q_{\rm max})^2
        \big)^2
        +
        4 \zeta^2 w^2
    } ,
\end{equation}
which expresses a balance between damage accumulation and healing. From this relation, the energy intensity required to attain a maximum level of damage $q_{\rm max}$ follows as
\begin{equation}\label{PAb5ph}
    \varepsilon(q_{\rm max})
    =
    \frac{\big(
            w^2
            -
            (1 - q_{\rm max})^2
        \big)^2
        +
        4 \zeta^2 w^2}{w^4}
        \frac{q_{\rm max}}{1-q_{\rm max}} .
\end{equation}
As expected, $\varepsilon(q_{\rm max})$ reduces to zero as $q_{\rm max} \to 0$ and diverges to infinity as $q_{\rm max} \to 1$. We also note that, by virtue of the existence of a steady state at $q_{\rm max}$, the integral in (\ref{tuS9Wr}) is well-defined and finite in the range $q_0 \leq q < q_{\rm max}$ and diverges to infinity at $q = q_{\rm max}$, indicating that the steady state is attained only asymptotically at infinite time.

\begin{figure}[h]
	\centering
	\begin{subfigure}[b]{0.45\textwidth}
		\includegraphics[width=0.99\textwidth]{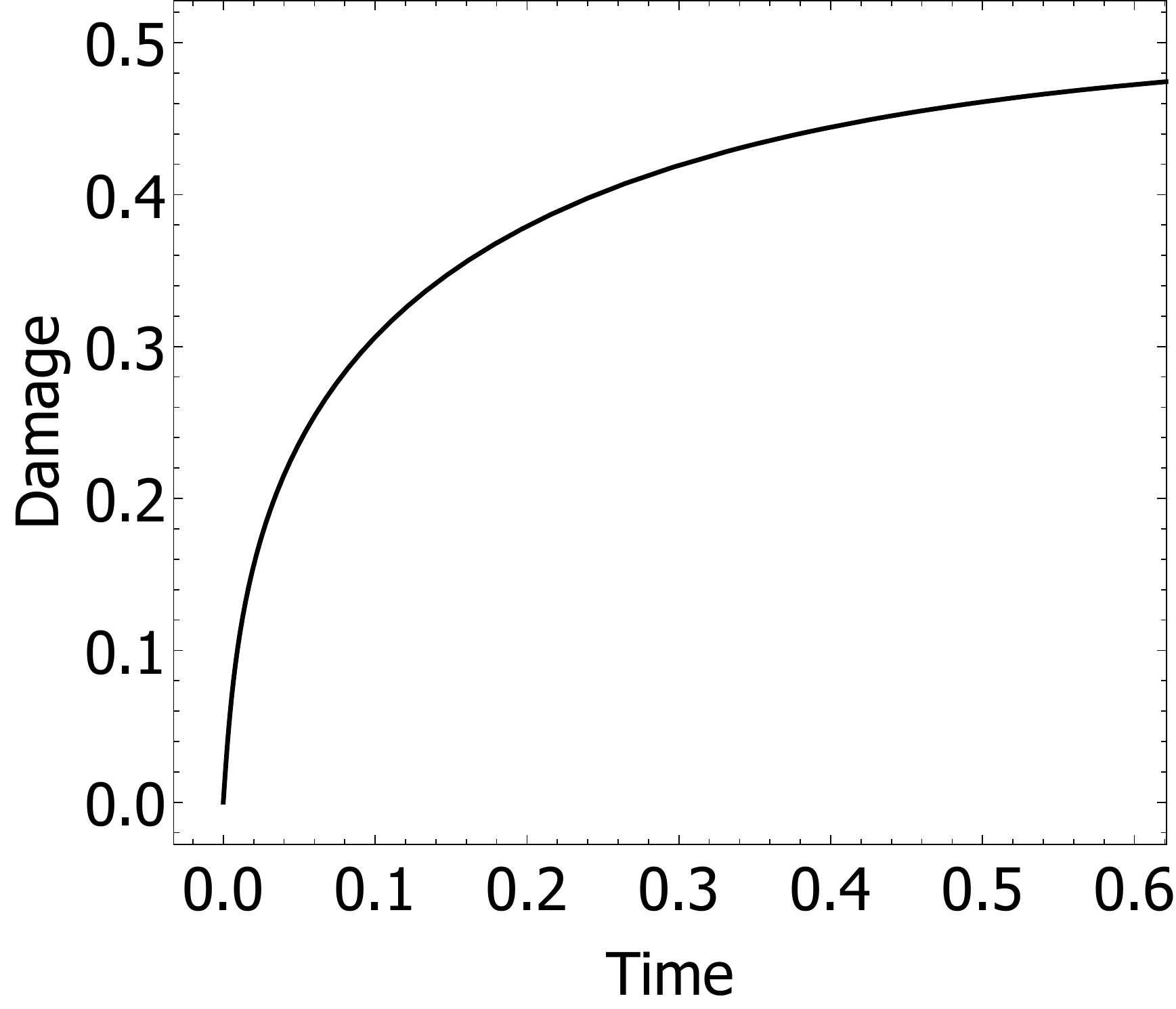}
		\caption{}
	\end{subfigure}
    $\quad$
	\begin{subfigure}[b]{0.45\textwidth}
		\includegraphics[width=0.99\textwidth]{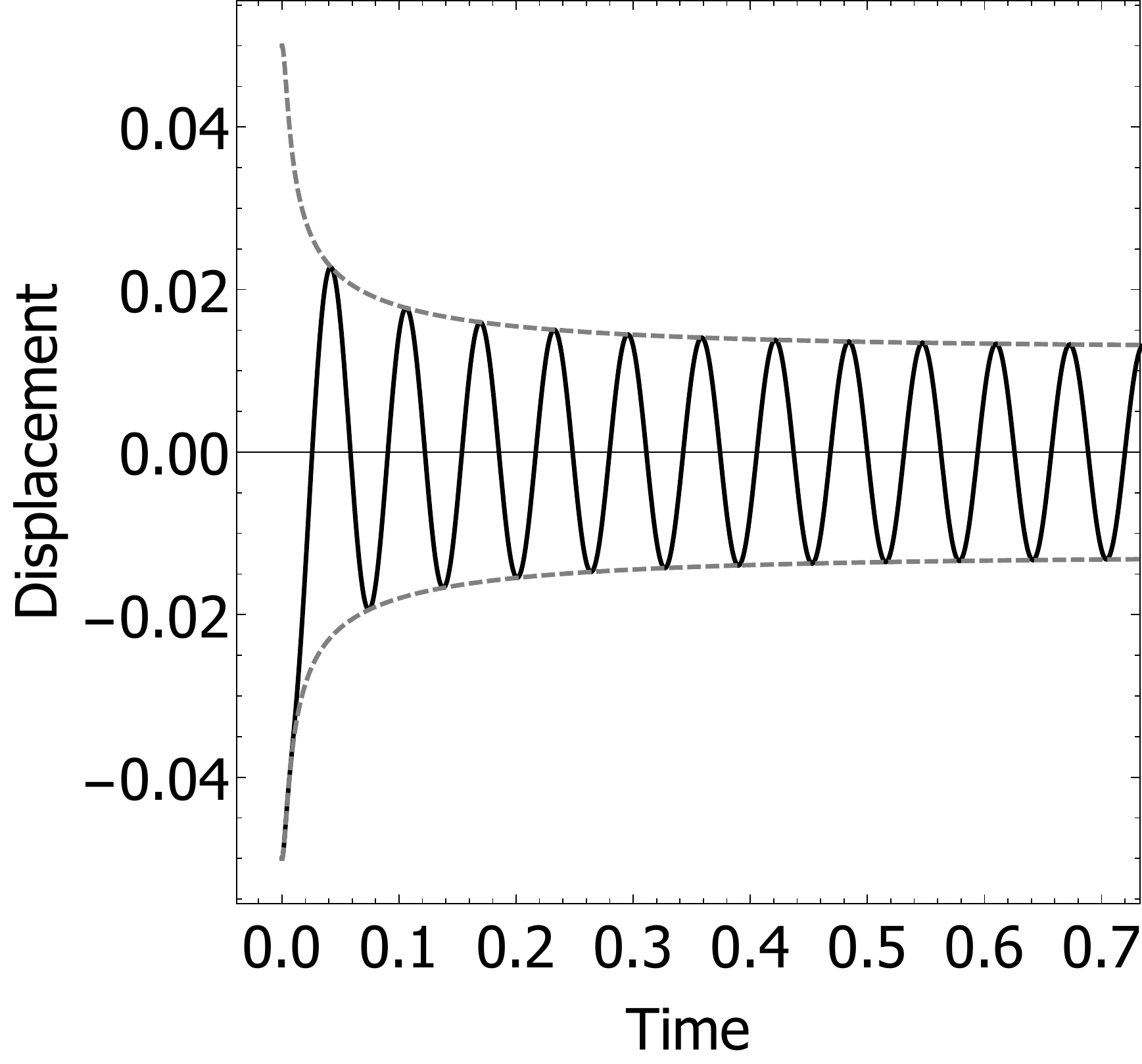}
		\caption{}
	\end{subfigure}
    \caption{Example of cell response to harmonic excitation. a) Damage state variable {\sl vs}.~time. b) Relative nucleus displacement and amplitude {\sl vs}.~time. Parameters: $t_r = 1$, $\omega=\omega_0=100$, $\zeta=1$, $q_{\rm max} = 1/2$. }\label{xeS7at}
\end{figure}

Fig.~\ref{xeS7at} shows an example of the WKB dynamics just elucidated for parameters: ${t_r} = 1$, $\omega=\omega_0=100$, $\zeta=1$, $q_{\rm max} = 1/2$. As may be seen from Fig.~\ref{xeS7at}, the state of damage of the cell evolves on the scale of the relaxation time ${t_r}$ for healing and tends asympotically to $q_{\rm max}$. The relative displacement of the nucleus is damped out on the shorter time scale $1/\zeta\omega_0$ and simultaneously amplified by the loss of stiffness due to damage on the time scale ${t_r}$. The competition between these two opposing effects results in a well-defined steady-state amplitude, which follows from (\ref{7tLYaD}) by taking the limit of $q(t) \to q_{\rm max}$. Correspondingly, the phase-space trajectory $(u(t),\dot{u}(t))$ converges to a stable limit cycle. The ability of WKB asymptotics to characterize the fast oscillations of the system and their slow modulation in time is remarkable.

During the off period, the governing equations (\ref{frEpi9}) reduce to
\begin{subequations}
\begin{align}
    &
    m \ddot{u}(t)
    +
    c \dot{u}(t)
    +
    (1 - q(t))^2 k u(t)
    =
    0 ,
    \\ & \label{SEQ3pl}
    {n} \dot{q}(t) + {d} q(t) = 0 .
\end{align}
\end{subequations}
Again, we assume that the duration $T_2$ of the off-period is much larger than the natural period of vibration $T_0$. Under these assumptions, in off-period we have
\begin{equation}
    u(t) = 0,
    \qquad
    q(t) = q_1 {\rm e}^{-(t-t_1)/{t_r}} ,
\end{equation}
outside a short transient decaying on the scale of $T_0$ immediately following $t_1$. Thus, modulo short transients during the off-period the cell is quiescent and repairs itself exponentially on the time scale of ${t_r}$.

\subsection{Fractional-step approximation of high-cycle limit}
Of special interest is the case in which the amount damage accumulated over each duty cycle is small. Thus, in the experiments of Mittelstein {\sl et al}.~\cite{Mittelstein:2019} the death of a significant fraction of the population requires the application of a large number of duty cycles of insonation. Correspondingly, the number of insonation pulses required to cause cell death is large, i.~e., $T/{t_r} \ll 1$. We proceed to obtain an effective equation describing the evolution of the system over larger numbers of duty cycles, or high-cycle limit. The effective equation follows by an appeal to the method of fractional steps \cite{Yanenko:1971}.

We recall that the duty cycle under consideration consists of an on-period of scaled duration $\tau_1 = T_1/{t_r}$ and an off-period of scaled duration $\tau_2 = T_2/{t_r}$. The entire scaled duration of the duty cycle is $\tau_1+\tau_2$. Assuming $\tau_1 \ll 1$, over a single on-period (\ref{r0biGU}) gives
\begin{equation}
    q_1
    \approx
    q_0
    +
    \tau_1
    \left(
        \dfrac
        {
            (1-q_0) w^4 \varepsilon
        }
        {
            \big(
                w^2
                -
                (1 - q_0)^2
            \big)^2
            +
            4 \zeta^2 w^2
        }
        -
        q_0
    \right) .
\end{equation}
Likewise, with $\tau_2 \ll 1$ over the subsequent off-period (\ref{SEQ3pl}) gives
\begin{equation}
    q_2
    \approx
    (1 - \tau_2) q_1 .
\end{equation}
Compounding the preceding relations and keeping terms of first order in $\tau_1$ and $\tau_2$ gives
\begin{equation}
    q_2
    \approx
    q_0
    +
    \tau_1
    \left(
        \dfrac
        {
            (1-q_0) w^4 \varepsilon
        }
        {
            \big(
                w^2
                -
                (1 - q_0)^2
            \big)^2
            +
            4 \zeta^2 w^2
        }
        -
        q_0
    \right)
    -
    \tau_2 q_0 .
\end{equation}
Rearranging terms gives the relation
\begin{equation}\label{w5iPrl}
    \frac{q_2 - q_0}{\tau_1+\tau_2}
    \approx
    \lambda
    \left(
        \dfrac
        {
            (1-q_0) w^4 \varepsilon
        }
        {
            \big(
                w^2
                -
                (1 - q_0)^2
            \big)^2
            +
            4 \zeta^2 w^2
        }
        -
        q_0
    \right)
    -
    (1-\lambda) q_0 ,
\end{equation}
where
\begin{equation}
    \lambda = \frac{\tau_1}{\tau_1 + \tau_2} ,
    \qquad
    1-\lambda = \frac{\tau_2}{\tau_1 + \tau_2} ,
\end{equation}
are the on-time fraction of the duty cycle, or duty factor, and the off-time fraction, respectively. Formally passing to the limit in (\ref{w5iPrl}) gives the differential equation
\begin{equation}\label{wlSw4w}
    \frac{dq}{d\tau}(\tau)
    +
    q(\tau)
    =
    \dfrac
    {
        \lambda (1-q(\tau)) w^4 \varepsilon
    }
    {
        \big(
            w^2
            -
            (1 - q(\tau))^2
        \big)^2
        +
        4 \zeta^2 w^2
    } ,
\end{equation}
which approximates slow damage evolution over larger numbers of duty cycles, or high-cycle limit. Again, the differential equation (\ref{wlSw4w}) is separable with solution
\begin{equation}\label{S3Apus}
    \tau
    =
    \int_0^q
    \frac
    {
        d\xi
    }
    {
        \dfrac
        {
            \lambda (1-\xi) w^4 \varepsilon
        }
        {
            \big(
                w^2
                -
                (1 - \xi)^2
            \big)^2
            +
            4 \zeta^2 w^2
        }
        -
        \xi
    } ,
\end{equation}
which is explicit up to a quadrature. As in the case of steady insonation, we note that the system attains a steady state at a maximum level of damage
\begin{equation}
    q_{\rm max}
    =
    \dfrac
    {
        \lambda (1-q_{\rm max}) w^4 \varepsilon
    }
    {
        \big(
            w^2
            -
            (1 - q_{\rm max})^2
        \big)^2
        +
        4 \zeta^2 w^2
    } ,
\end{equation}
at which point damage accumulation and healing balance each other. The energy intensity required to attain a maximum level of damage $q_{\rm max}$ follows as
\begin{equation}\label{qaCHu8}
    \varepsilon(q_{\rm max}, \lambda)
    =
    \frac{\big(
            w^2
            -
            (1 - q_{\rm max})^2
        \big)^2
        +
        4 \zeta^2 w^2}{\lambda w^4}
        \frac{q_{\rm max}}{1-q_{\rm max}} .
\end{equation}
As expected, $\varepsilon(q_{\rm max}, \lambda)$ reduces to zero as $\lambda \to 0$ and reduces to (\ref{PAb5ph}) for $\lambda = 1$. We also note that the integral in (\ref{S3Apus}) is well-defined and finite in the range $q_0 \leq q < q_{\rm max}$ and diverges to infinity at $q = q_{\rm max}$, indicating that the steady state is attained only asymptotically.

\begin{figure}[h]
	\centering
	\begin{subfigure}[b]{0.45\textwidth}
		\includegraphics[width=0.99\textwidth]{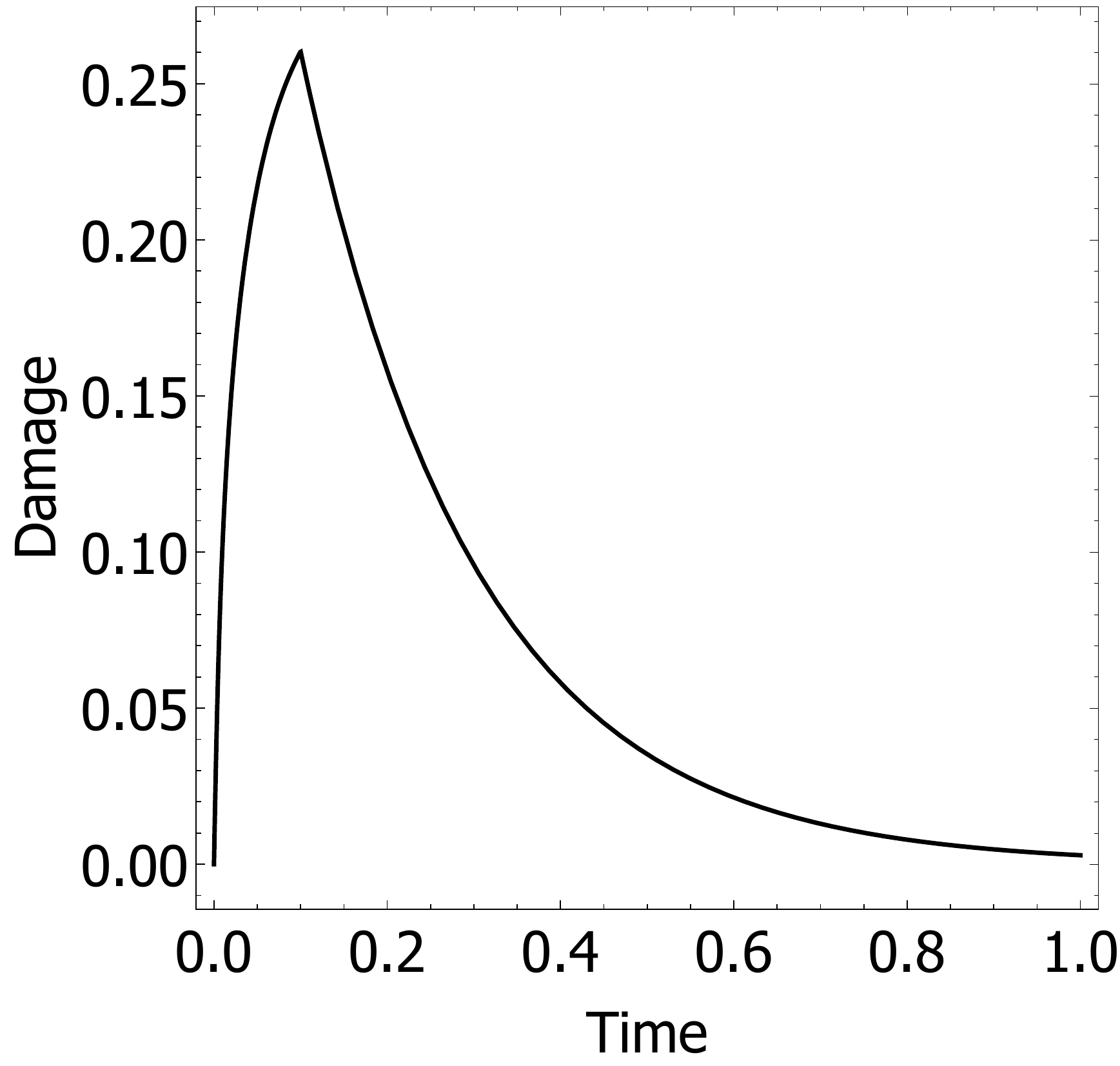}
		\caption{}
	\end{subfigure}
    $\quad$
	\begin{subfigure}[b]{0.45\textwidth}
		\includegraphics[width=0.99\textwidth]{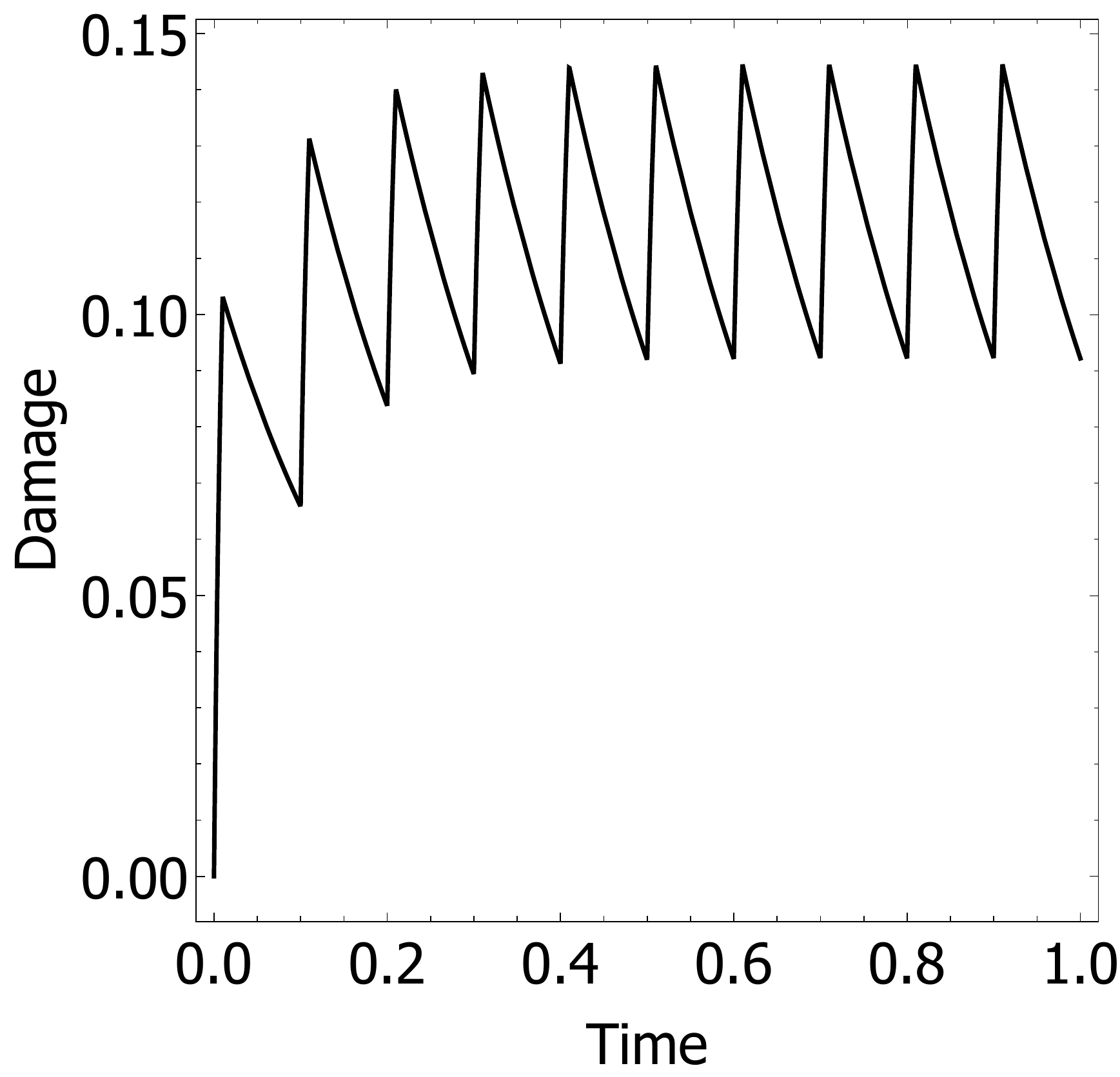}
		\caption{}
	\end{subfigure}
	\begin{subfigure}[b]{0.45\textwidth}
		\includegraphics[width=0.99\textwidth]{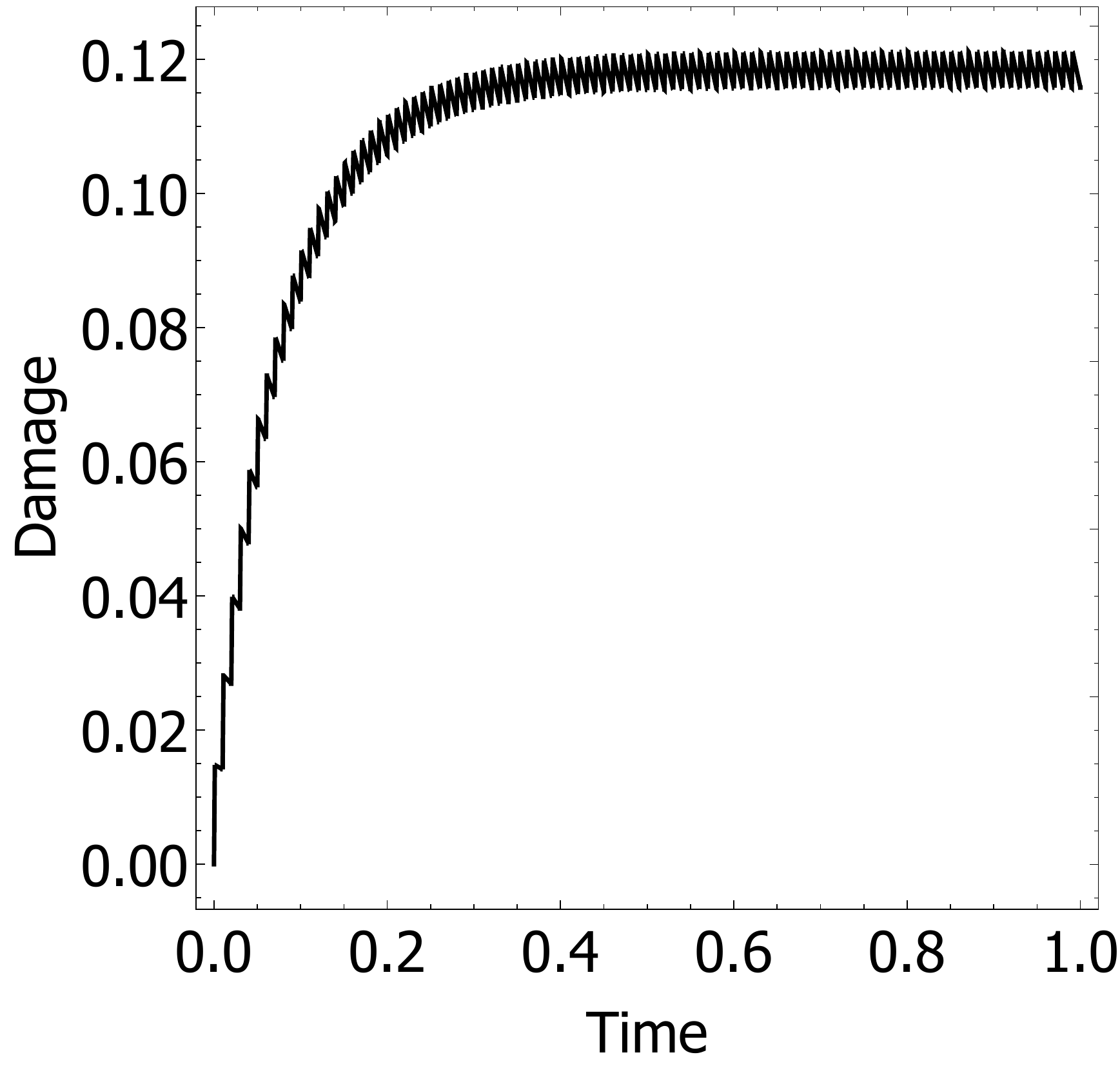}
		\caption{}
	\end{subfigure}
    $\quad$
	\begin{subfigure}[b]{0.45\textwidth}
		\includegraphics[width=0.99\textwidth]{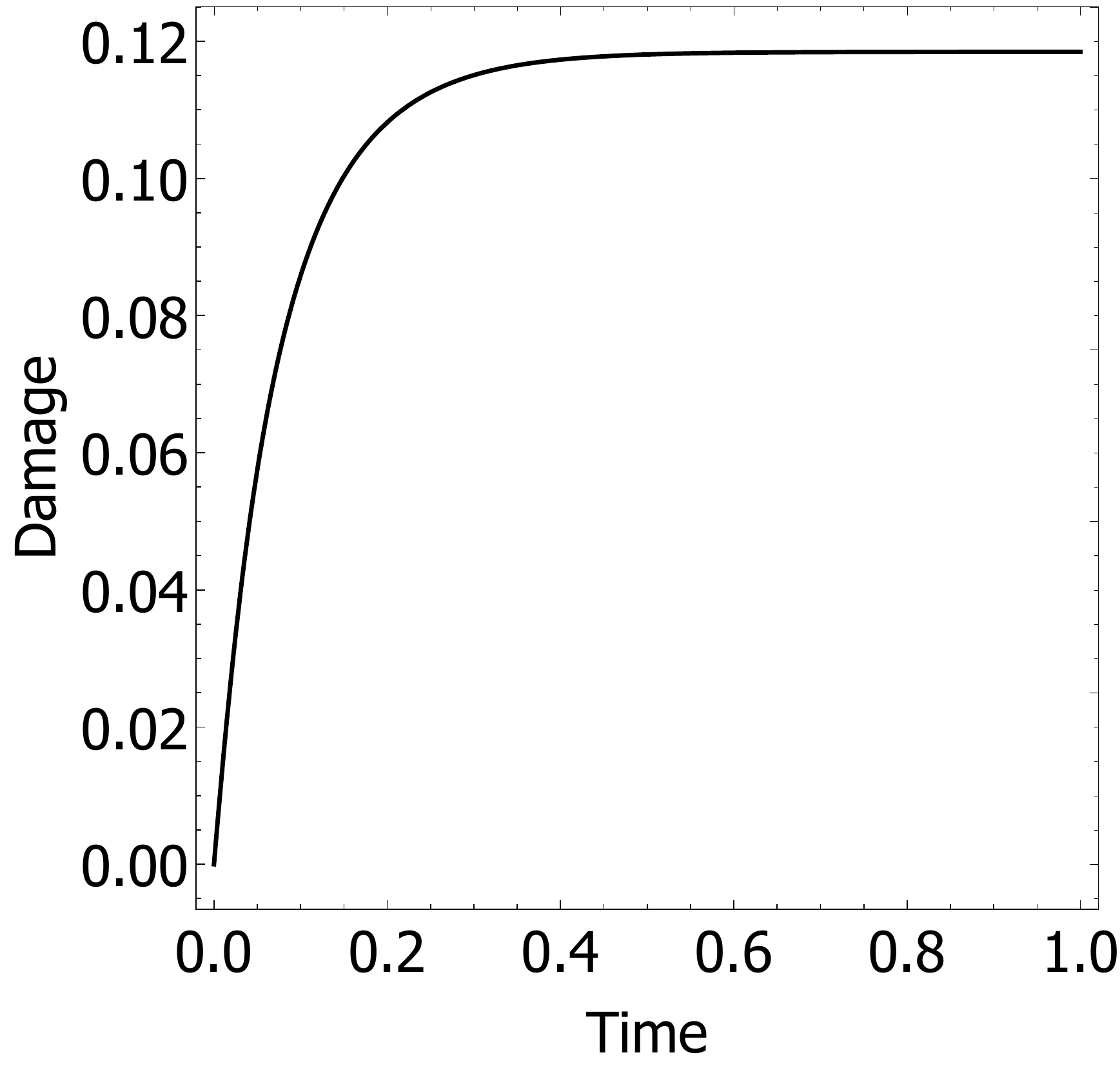}
		\caption{}
	\end{subfigure}
    \caption{Convergence of the damage evolution to the high-cycle limit as pulse repetition period $T$ becomes much smaller than the characteristic time ${t_r}$ for healing, cf.~Fig.~\ref{8w5RPa}a (inset). Parameters: ${t_r} = 10$, $\lambda=1/10$, $\omega=\omega_0=100$, $\zeta=1/10$, $q_{\rm max} = 1/2$. a) $T = 1$. b) $T=1/10$. c) $T=1/100$. d) Damage evolution predicted by the high-cycle limit equation (\ref{wlSw4w}). }\label{pu3ruC}
\end{figure}

The convergence of the damage evolution to the high-cycle limit as the pulse repetition period $T$ becomes much smaller than the characteristic time ${t_r}$ for healing is illustrated in Fig.~\ref{pu3ruC}, which corresponds to the choice of parameters: ${t_r} = 10$, $\lambda=1/10$, $\omega=\omega_0=100$, $\zeta=1/10$, $q_{\rm max} = 1/2$. Figs.~\ref{pu3ruC}a-c show the evolution of the damage state variable obtained by solving directly the WKB eq.~(\ref{r0biGU}) and eq.~(\ref{SEQ3pl}) for $T = 1$, $1/10$ and $1/100$, respectively. As expected, damage accumulates during the off-period and otherwise relaxes at all times, resulting in a characteristic saw-tooth profile. Fig.~\ref{pu3ruC}d shows the corresponding evolution of the damage state variable predicted by the effective fractional-step eq.~(\ref{wlSw4w}). Evidently, the high-cycle limiting curve is smooth and represents a weak limit of the damage evolution curves as the number of duty cycles tends to infinity, respectively, the pulse duration cycle tends to zero.

\subsection{Cell death}

We recall that the state variable $q(t)$ measures the amount of damage sustained by a cell at time $t$. A plausible assumption is that a cell becomes unviable and dies when $q(t)$ attains a critical value $q_c$. In light of our previous discussion, this condition cannot be met if $q_{\rm max} \leq q_c$, i.~e., if the maximum accumulated damage induced by insonation is less that the critical value. Conversely, it follows from (\ref{qaCHu8}),  that cell death requires a minimum level of energy deposition
\begin{equation}\label{jihuS4}
    \varepsilon \geq \varepsilon(q_c, \lambda) .
\end{equation}
If this condition is met, then in the high-cycle limit the time-to-death of a cell follows from (\ref{S3Apus}) as
\begin{equation}\label{qApru3}
    \tau_c
    =
    \int_0^{q_c}
    \frac
    {
        d\xi
    }
    {
        \dfrac
        {
            \lambda (1-\xi) w^4 \varepsilon
        }
        {
            \big(
                w^2
                -
                (1 - \xi)^2
            \big)^2
            +
            4 \zeta^2 w^2
        }
        -
        \xi
    } ,
\end{equation}
otherwise $\tau_c = +\infty$ and the cell survives for all time. The corresponding number of insonation pulses is
\begin{equation}
    N_c = \frac{{n}}{{d}} \frac{\tau_c}{T} ,
\end{equation}
were $T$ is the total pulse duration.

As noted in the introduction, this type of system failure by slow damage accumulation over many cycles is observed in other systems, notably inert structural materials, in which context it is known as {\sl high-cycle mechanical fatigue} \cite{Suresh:1998}. The number of loading cycles to failure is correspondingly known as the {\sl fatigue life} of the material. In this analogy, cell death by slow damage accumulation over many cycles may be thought of as a form of {\sl mechanical cell fatigue}, and the number of cycles $N_c$ to death as the {\sl fatigue life} of the cell.

\subsection{Variability within a cell population}\label{dRA7Rl}

A typical population of cancerous cells exhibits broad variation in geometry and mechanical properties. This variability is strongly suggested by the cell-death curves observed by Mittelstein {\sl et al.} \cite{Mittelstein:2019}, which show that some cells die much earlier than others. In order to capture this gradual cell necrosis, we regard the parameters governing the evolution of the cells as random and a cell population as a sample drawn from the probability distribution of the parameters. By virtue of the variability of the sample, parts of the population have a relatively short time-to-death and die early, whereas other parts have a comparatively longer time-to-death and die later, resulting in the gradual estimated cell death curves observed experimentally, Fig.~\ref{1HYBrW}.

The statistics of the time-to-death can be estimated simply by means of a linear sensitivity analysis, cf., e.~g., \cite{Sullivan:2015}. We see from (\ref{qApru3}) that the time-to-death $t_c = \tau_c t_r$ depends on the cell parameters $(t_r, \omega_0, \zeta, q_c)$, respectively, the relaxation time for healing, the natural frequency of vibration and the damping ratio; and on the process parameters $(\varepsilon, \omega, \lambda)$, respectively, the energy intensity, frequency and on-period fraction of the insonation. For simplicity, we assume that the process parameters can be controlled exactly and are uncertainty-free. Contrariwise, the cell parameters define a multivariate random variable $X \equiv (t_r, \omega_0, \zeta, q_c)$, with probability distribution reflecting the variability of the cell population.

Owing to the randomness of the cell population, the time-to-death $t_c$ itself defines a random variable $Y$. In terms of these random variables, (\ref{qApru3}) defines a relation of the form
\begin{equation}\label{c6ataP}
    Y = f(X) .
\end{equation}
In order to estimate the variability in the time-to-death random variable $Y$, we make a small-deviation approximation
\begin{equation}
    Y \approx f(\bar{X}) + Df(\bar{X}) (X-\bar{X}) + h.o.t.,
\end{equation}
where
\begin{equation}
    \bar{X}
    =
    \mathbb{E}(X)
    \equiv
    (\bar{t}_r , \bar{\omega}_0, \bar{\zeta}, \bar{q}_c)
\end{equation}
is the mean value of the cell parameters and $Df(\bar{X})$ are sensitivity parameters. The average time-to-death then follows as
\begin{equation}\label{4Abudl}
    \bar{Y}
    =
    \mathbb{E}(Y)
    \approx
    f(\bar{X}) + h.o.t.
\end{equation}
In addition, a measure of the variability of $Y$ is given by the variance
\begin{equation}\label{vowE5U}
    \sigma_Y^2
    =
    \mathbb{E}((Y-\bar{Y})^2)
    =
    Df(\bar{X})^T
    \mathbb{E} ((X-\bar{X}) \otimes (X-\bar{X}))
    Df(\bar{X})
    =
    Df(\bar{X})^T \Sigma Df(\bar{X}) ,
\end{equation}
where
\begin{equation}
    \Sigma
    =
    \mathbb{E}((X-\bar{X}) \otimes (X-\bar{X}))
\end{equation}
is the covariance matrix of the cell parameters.

We note that, for small deviations, the mean time-to-death of the cell population is obtained by evaluating (\ref{qApru3}) at the mean value $\bar{X} = (\bar{t}_r, \bar{\omega}_0, \bar{\zeta}, \bar{q}_c)$ of the cell parameters, cf.~eq.~(\ref{4Abudl}), with the result
\begin{equation}
    \bar{t}_c
    =
    \bar{t}_r
    \int_0^{\bar{q}_c}
    \frac
    {
        d\xi
    }
    {
        \dfrac
        {
            \lambda (1-\xi) \bar{w}^4 \varepsilon
        }
        {
            \big(
                \bar{w}^2
                -
                (1 - \xi)^2
            \big)^2
            +
            4 \bar{\zeta}^2 \bar{w}^2
        }
        -
        \xi
    } ,
\end{equation}
where we write $\bar{w} = \omega/\bar{\omega}_0$ and we assume that (\ref{jihuS4}) is satisfied with $q_c = \bar{q}_c$.  Likewise, the requisite sensitivity parameters $Df(\bar{X})$ follow by differentiating (\ref{4Abudl}) with respect to the cell parameters and evaluating the resulting integrals at their mean value.

\begin{figure}[h]
	\centering
	\begin{subfigure}[b]{0.45\textwidth}
		\includegraphics[width=0.99\textwidth]{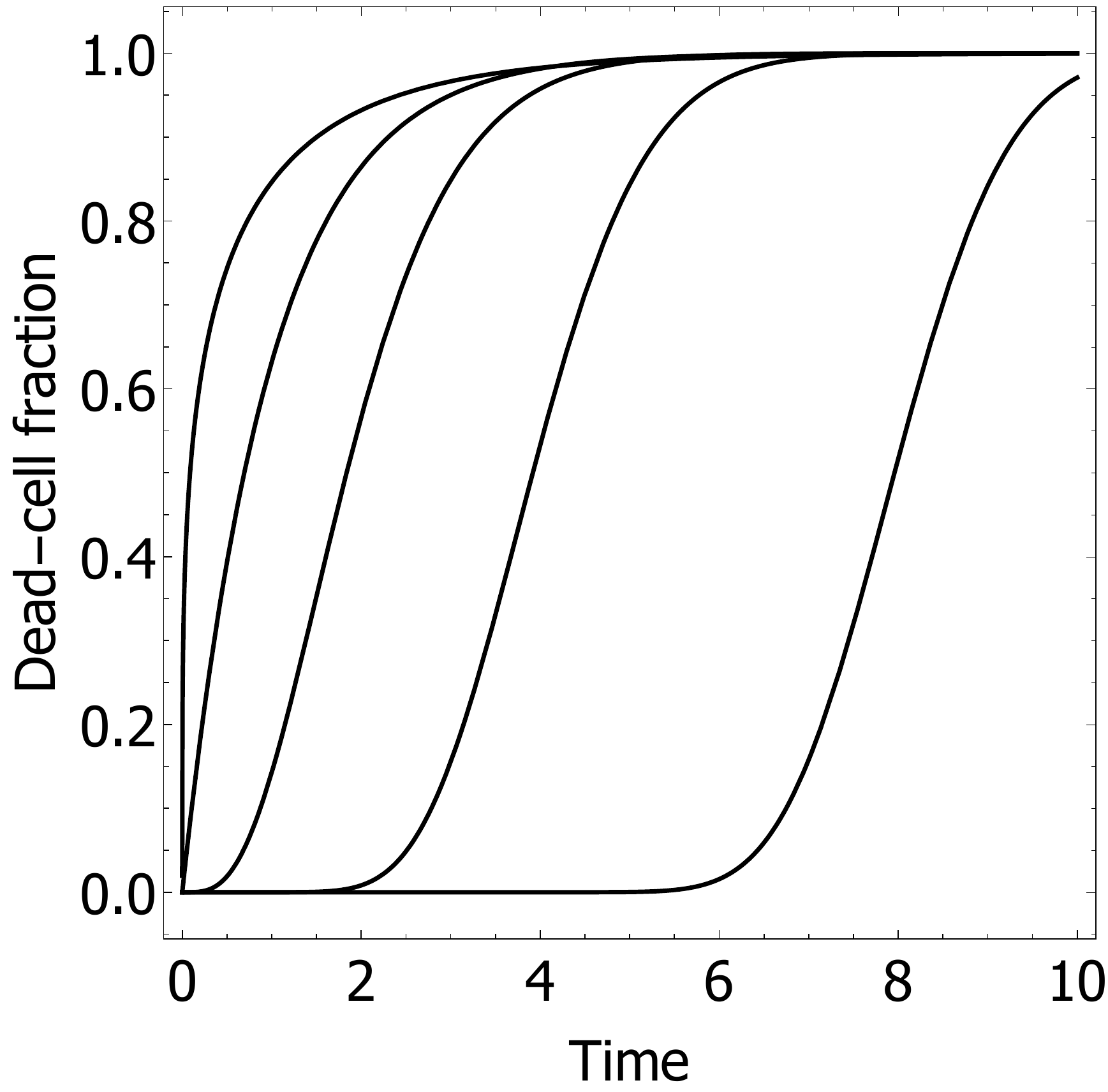}
		\caption{}
	\end{subfigure}
    $\quad$
	\begin{subfigure}[b]{0.45\textwidth}
		\includegraphics[width=0.99\textwidth]{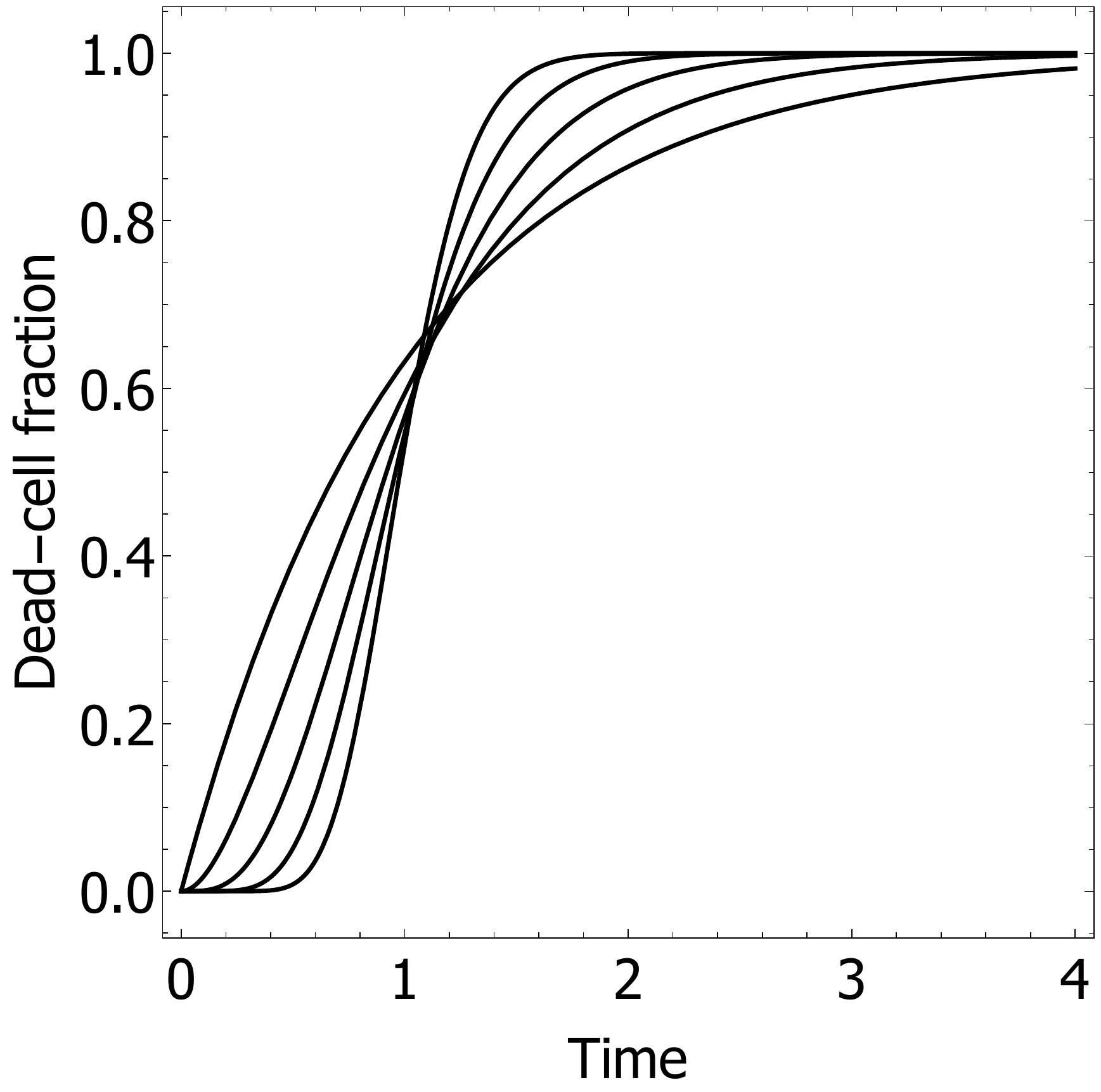}
		\caption{}
	\end{subfigure}
    \caption{Dead-cell fraction {\sl vs}.~time curves obtained from the Gamma-distribution. a) $\sigma_Y^2=1$, $\bar{t}_c$ $=$ $1/2$, $1$, $2$, $4$ and $8$. b) $\bar{t}_c=1$, $\sigma_Y^2$ $=$ $1$, $1/2$, $1/4$, $1/8$ and $1/16$.}\label{9uDRuc}
\end{figure}

Simple forms of the probability distribution of $t_c$ are fully determined by the statistics $\bar{Y}$ and $\sigma_Y^2$. For instance, if we hypothesize a gamma distribution
\begin{equation}\label{jL8Uth}
    p(Y)
    =
    \frac{1}{\Gamma(k) \theta^k} Y^{k-1} {\rm e}^{-Y/\theta} ,
\end{equation}
then the parameters of the distribution follow as
\begin{equation}
    \bar{Y} = k \theta,
    \qquad
    \sigma_Y^2 = k \theta^2 .
\end{equation}
The fraction of the cell population with a time-to-death less or equal to $t$ is given by the cumulative distribution function
\begin{equation}\label{Pru4oW}
    F(t) = P(Y \leq t) .
\end{equation}
For the gamma distribution (\ref{jL8Uth}), we have
\begin{equation}\label{pAt8ow}
    F(t)
    =
    1-\dfrac{\Gamma \left(k, t/\theta\right)}{\Gamma (k)} ,
\end{equation}
where $\Gamma$ is the gamma function. The resulting dead-cell fraction {\sl vs}. time curves are illustrated in Fig.~\ref{9uDRuc}.

\begin{figure}[!h]
	\centering
	\includegraphics[width=0.5\textwidth]{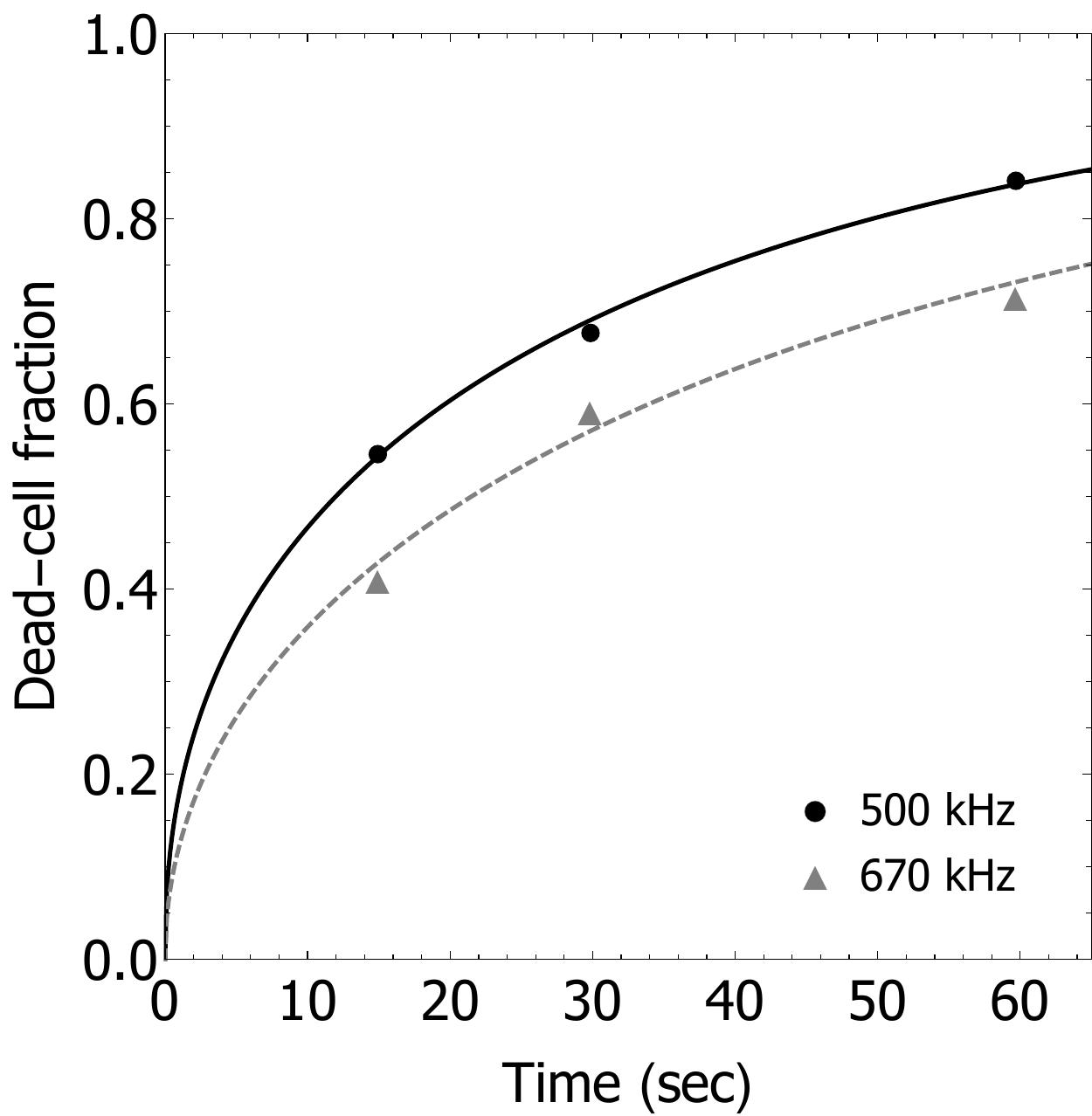}
	\caption{$\Gamma$-distribution fit of cell-death time data \cite{Mittelstein:2019} for cell line K-562 at focal pressure 1.4 MPa, pulse duration 100ms, 10\% duty cycle at two insonation frequencies 500 kHz and 670 kHz.}.
	\label{fig:celldeathfit}
\end{figure}

\begin{table}[h]
	\centering
	\begin{tabular}{|c|c|c|c|}
        \hline
        \multicolumn{2}{|c|}{500 kHz}  &
        \multicolumn{2}{|c|}{670 kHz} \\
		\hline
		$\bar{Y}$ (sec) & $\sigma_Y$ (sec) & $\bar{Y}$ (sec) & $\sigma_Y$ (sec) \\
		\hline
		30.5 & 46.6 & 49.4 & 71.36 \\
		\hline
	\end{tabular}
	\caption{Mean and standard deviation obtained by fitting to cell-death time data \cite{Mittelstein:2019} for cell line K-562 at focal pressure 1.4 MPa, pulse duration 100ms, 10\% duty cycle at two insonation frequencies 500 kHz and 670 kHz. }
	\label{tab:expparams}
\end{table}

By way of example, Fig.~\ref{fig:celldeathfit} shows a least-squares fit of the cell-death time data of \cite{Mittelstein:2019} using the function $F(t)$ obtained from the $\Gamma$ distribution, eq.~(\ref{pAt8ow}). The data corresponds to the cell line K-562 at focal pressure 1.4 MPa, pulse duration 100 ms and 10\% duty cycle. The mean and standard deviation derived from the fit are listed in Table~\ref{tab:expparams}. As may be seen from the figure, the $\Gamma$ distribution provides an adequate fit of the data.

\section{Comparison with experiment}

We proceed to assess the ability of the proposed dynamical model to account for the experimentally observed trends summarized in Section~\ref{i9UQnX}.

\subsection{Qualitative comparison}

\begin{figure}[h]
	\centering
	\begin{subfigure}[b]{0.45\textwidth}
		\includegraphics[width=0.99\textwidth]{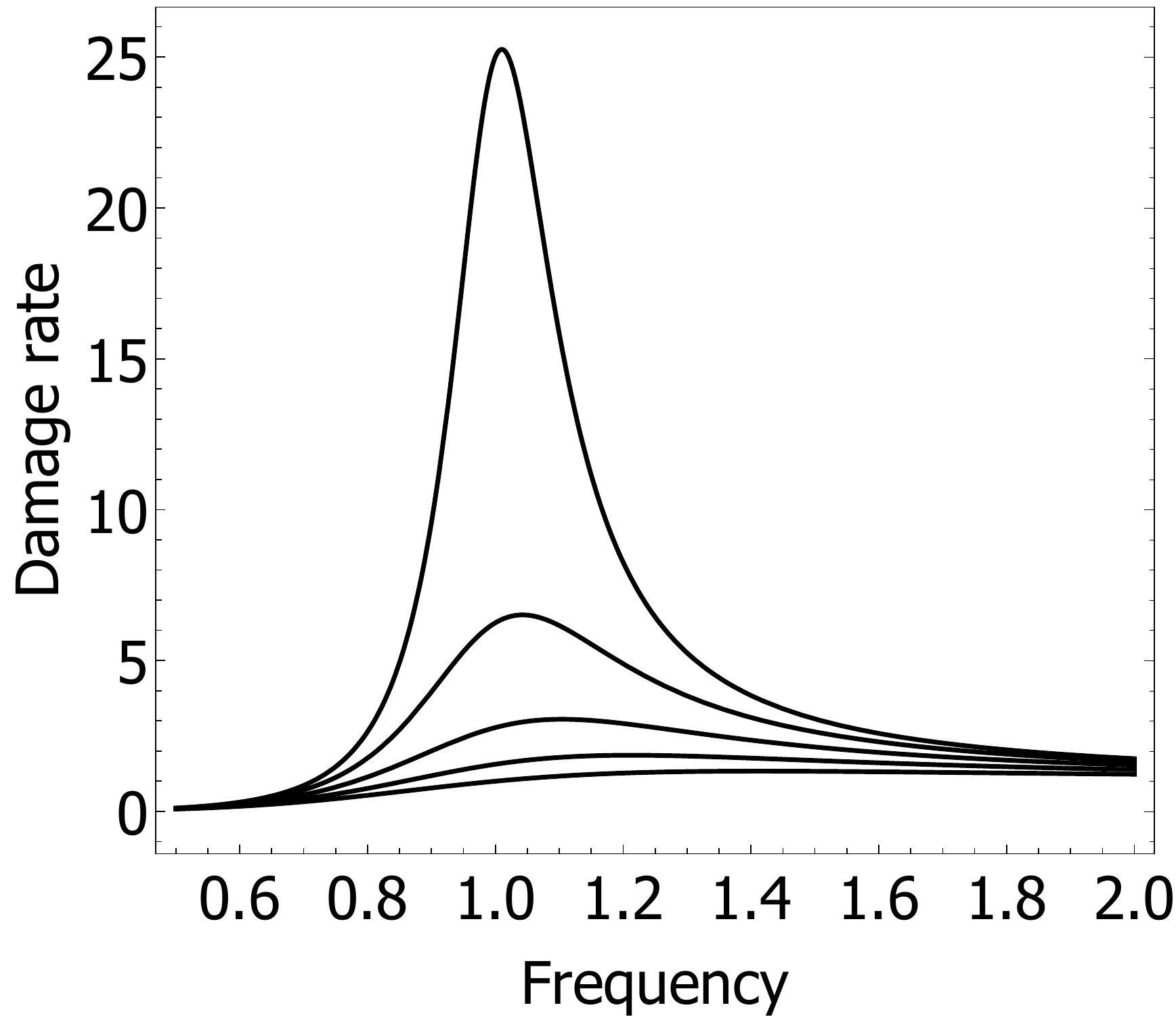}
		\caption{}
	\end{subfigure}
    $\quad$
	\begin{subfigure}[b]{0.45\textwidth}
		\includegraphics[width=0.99\textwidth]{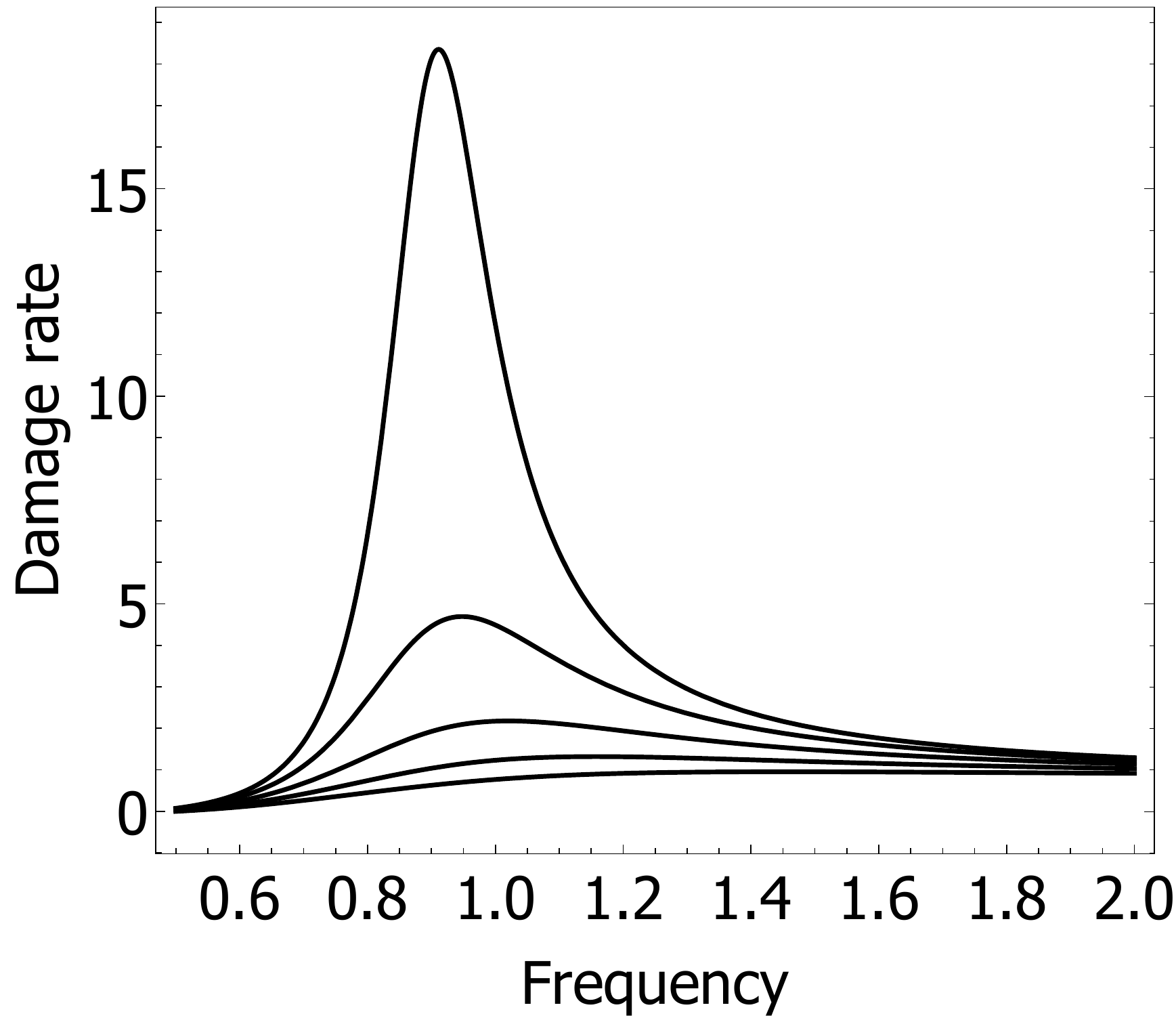}
		\caption{}
	\end{subfigure}
    \caption{Damage accumulation rate as a function of insonation frequency. Parameters: $\omega_0 = 1$, $\varepsilon = 1$, $\lambda = 1$, $t_r = 1$, $\zeta = 1/10$, $2/10$, $3/10$, $4/10$, $5/10$. a) Pristine cell, $q=0$. b) Damaged cell, $q=1/10$.}\label{w4UDrA}
\end{figure}

We note that the experimentally observed dead-cell fraction vs.~time curves exhibit the sigmoidal form predicted by the proposed dynamical model, cf.~Figs.~\ref{1HYBrW} and \ref{9uDRuc}, which can be used to fit the experimental curves. More importantly, the model {\sl explains} the observed dead-cell fraction curves as a result of cell-to-cell variability, specifically, the random distribution of times-to-death in the cell population. Furthermore, the time-to-death of an individual cell is predicted by the model explicitly as a function of cell parameters $(t_r, \omega_0, \zeta, q_c)$ and process parameters $(\varepsilon, \omega, \lambda)$, e.~g., through eq.~(\ref{qApru3}) in the high-cycle limit. Owing to the variability of the cell population, the cell parameters may be assumed to be random and, by an appeal to linear sensitivity analysis, the mean and variance of the cell time-to-death can be related to the mean values and covariance matrix of the cell parameters, eqs.~(\ref{qApru3}) and (\ref{vowE5U}). Thus, if the statistics of the cell parameters is known, the time-to-death statistics and, correspondingly, the dead-cell fraction curves, are given explicitly by the model. In this manner, the model relates the observed dead-cell fraction curves to fundamental mechanical properties of the cell such as mass, stiffness, viscosity and damage tolerance.

The dynamical model also predicts the dependence of the dead-cell fraction curves on pulse duration observed experimentally, Fig.~\ref{v0EZTl}. Indeed, this trend is exhibited by the damage evolution curves shown in Fig.~\ref{pu3ruC}. A careful inspection of these curves shows that the maximum level of damage attained within the insonation cycles decreases as the pulse duration decreases relative to the characteristic time for healing. Thus, for long pulses the cells have time to accumulate large amounts of damage during the on-period of the pulse. For shorter pulses, the extent of damage accumulation is comparatively less. If the pulse duration is comparable to---or smaller than---the relaxation time for healing, the cell does not have sufficient time to recover during the off-period of the cycle, and the trend persists over repeated cycles. Therefore, according to the model the dependence of the dead-cell fraction curves on pulse duration is the result of a delicate interplay between the pulse repetition period and pulse duration, the cell dynamics, which determines the rate at which damage accumulates and the kinetics of cell healing, which determines the rate at which damage is restored.

The dynamical model also exhibits the oncotripsy effect, i.~e., the insonation-frequency dependence of the cell response and the window of opportunity for selective cell ablation. Fig.~\ref{w4UDrA} shows the damage accumulation rate $\dot{q}$ computed from (\ref{wlSw4w}) as a function of insonation frequency, damping ratio and state of damage. The parameters used in the figure are: $\omega_0 = 1$, $\varepsilon = 1$, $\lambda = 1$, $t_r = 1$, $\zeta = 1/10$, $2/10$, $3/10$, $4/10$, $5/10$, $q = 0$, $1/10$. As may be seen from the figure, the damage rate peaks sharply in the vicinity of the undamped resonant frequency $\omega = \omega_0$. The damage accumulation rate is largest for a pristine cell, $q=0$, and persists, albeit somewhat reduced, after the cell sustains damage, $q=1/10$. This frequency dependence is clearly apparent in the experimental data, Fig.~\ref{1HYBrW}b.

%%%%%%%%%%%%%%%%%%%%%%%
\subsection{Quantitative comparison}
%%%%%%%%%%%%%%%%%%%%%%%

A quantitative comparison between the predicted cell death times and experimental data provides a measure of validation of the model. We recall that the death time $t_r$ of a cell characterized by parameters $X \equiv (t_r, \omega_0, \zeta, q_c)$ is given analytically by (\ref{qApru3}). We regard $X$ as a multivariate random variable with a certain probability distribution reflecting the variability of the cell population. Owing to this variability, the time-to-death $t_c$ itself defines a random variable $Y$, in terms of which (\ref{qApru3}) is to be regarded as a response function of the form (\ref{c6ataP}).

In order to exercise the linearized sensitivity framework formulated in Section~\ref{dRA7Rl}, we need to know the average values $\bar{X}$ of the parameters $X$ for a given cell population and their covariance matrix $\Sigma$. In lieu of direct characterization, we estimate these statistics as follows. We begin by assuming that the parameters $X$ are independent and log-normal distributed with unknown mean $\bar{X}$ and diagonal covariance matrix $\Sigma$. From this distribution, we generate a random sample $\{X_i,\ i=1,\dots,N\}$ of size $N=1000$, compute the corresponding cell-death times $\{t_i,\ i=1,\dots,N\}$ using (\ref{qApru3}) and evaluate the fraction of the cell population with a time-to-death less or equal to $t$ as, cf.~eq.~(\ref{Pru4oW}),
\begin{equation}
    F(t) = \frac{1}{N} \#\{t_i \leq t,\ i=1,\dots, N \}, % P(Y \leq t) .
\end{equation}
where $\#$ is the counting measure. The statistics $\bar{X}$ and $\Sigma$ are then obtained by means of a least-square fit to the data. The results are listed in Table~\ref{tab:randomvariable}.

\begin{table}[h]
    \centering
    \begin{tabular}{|l|c|c|c|c|c|}
                   \hline
                   &$t_r$ (sec) & $\omega_0$ (rads/sec) & $\zeta$ & $q_c$ \\ \hline
                   Mean & 100 & 3142 & 0.7 & 0.136 \\ \hline
                   Standard deviation & 10 & 393 & 0.175 & 0.0136 \\ \hline
    \end{tabular}
    \caption{Estimated mean, standard deviation and sensitivities of cell parameters.}
    \label{tab:randomvariable}
\end{table}

\begin{figure}[!h]
	\centering
	\begin{subfigure}[t]{0.45\textwidth}
		\includegraphics[width=1\textwidth]{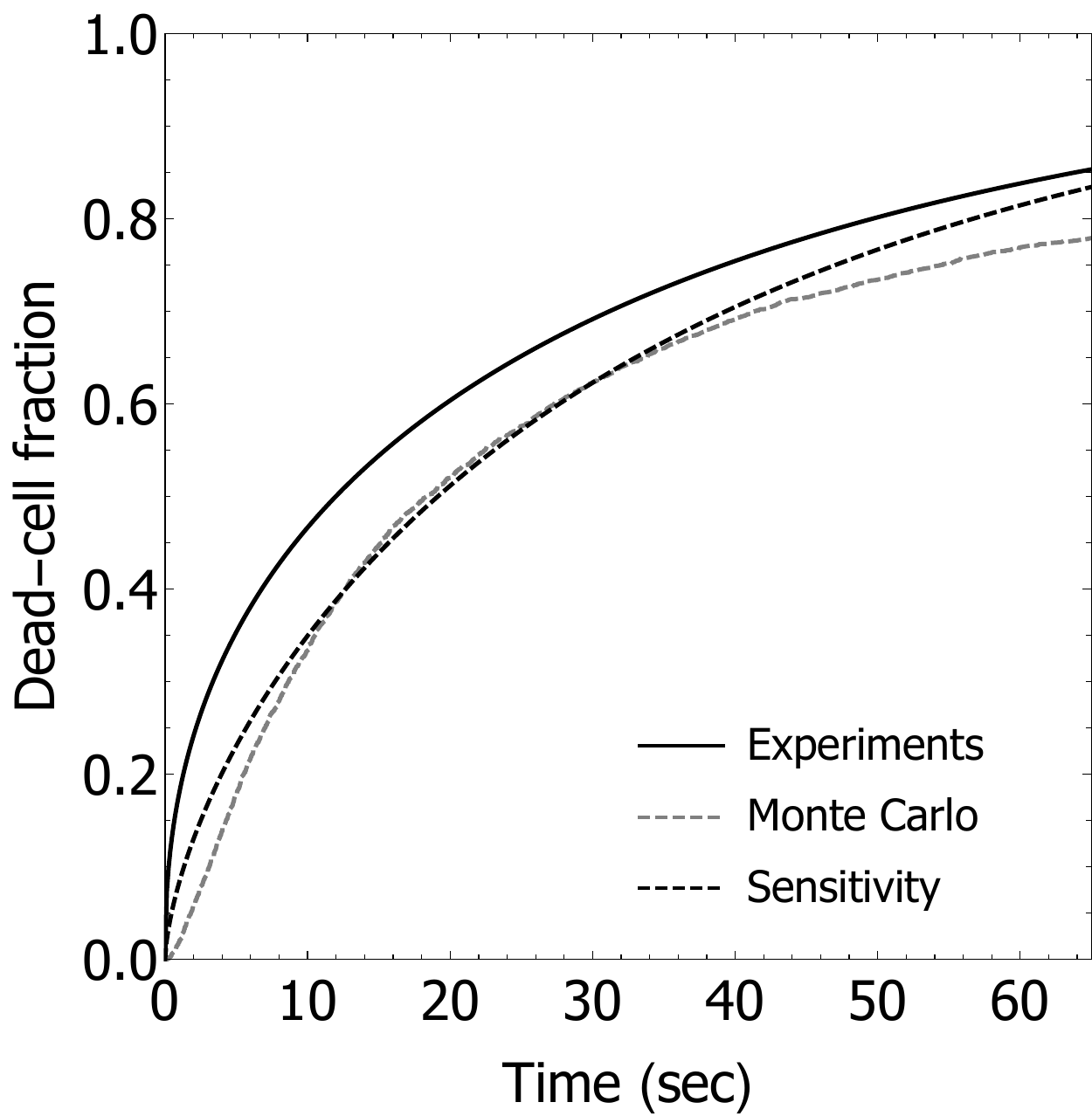}
	\end{subfigure}
	\hfill
	\begin{subfigure}[t]{0.45\textwidth}
		\includegraphics[width=1\textwidth]{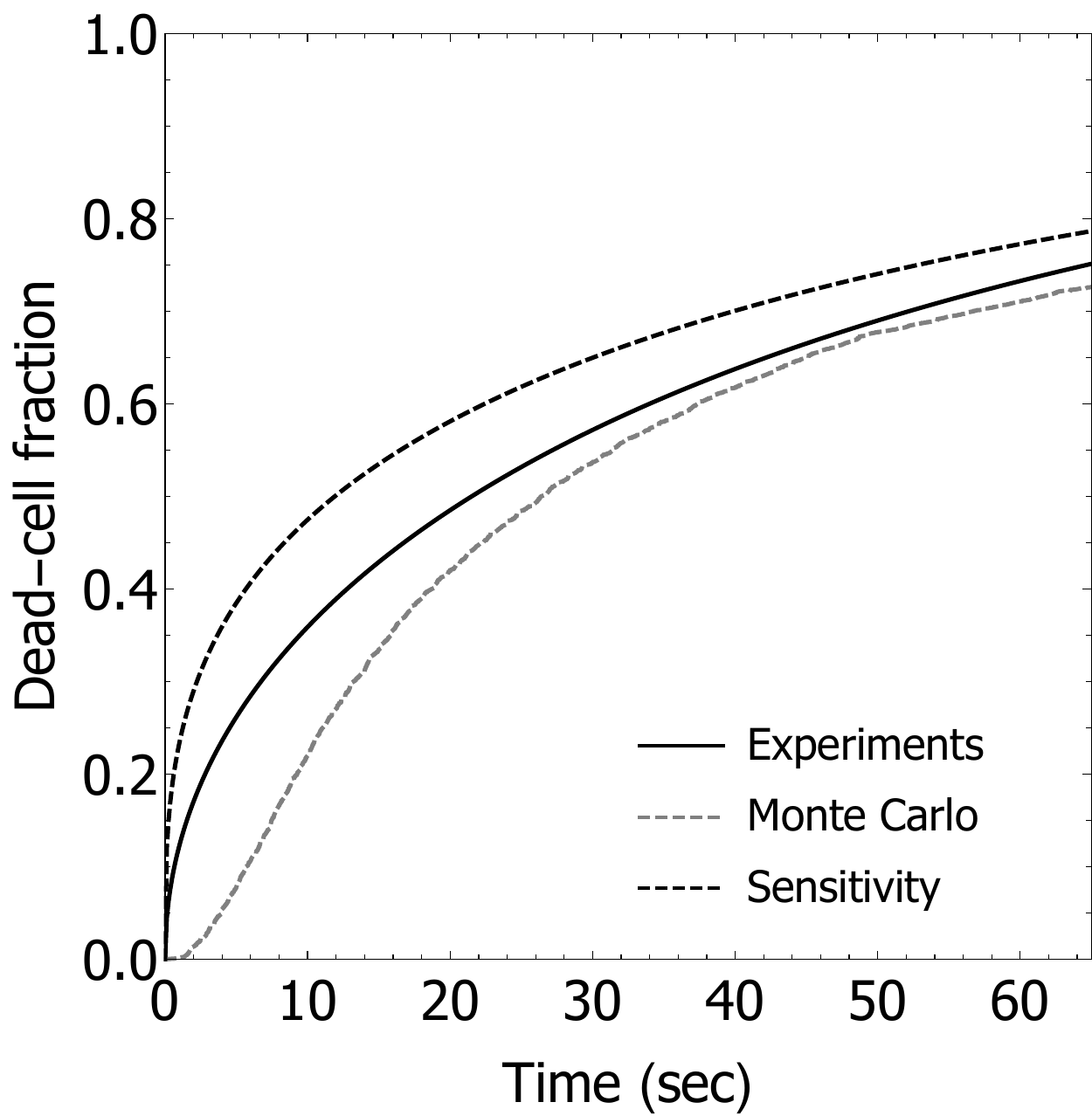}
	\end{subfigure}
    \caption{Comparison of predicted cell-death fraction with experimental data from \cite{Mittelstein:2019} for a focal pressure of 1.4MPa, pulse duration 100ms, duty cycle 10\% and frequencies 500kHz and 670kHz. The experimental data is represented through the $\Gamma$-distribution fit shown in Fig.~\ref{fig:celldeathfit}.} \label{fig:ECD_ExpVsFEM}
\end{figure}

Finally, we are in a position to compare predicted cell-death curves with the experimental data. Fig.~\ref{fig:ECD_ExpVsFEM} shows computed cell-death curves for log-normal independent cell population parameters, with mean values and standard deviations as in Table~\ref{tab:randomvariable}, together with experimental data from \cite{Mittelstein:2019}. The predicted curves are computed directly via Monte Carlo based on a sample of size $N=1000$ and by means of the linearized-sensitivity approximation. As may be seen from the figure, the linearized-sensitivity curve closely approximates the Monte Carlo curve, which establishes the validity of the linearized-sensitivity approximation under the conditions of the experiments. In addition, both the linearized-sensitivity and the Monte Carlo curves match closely the experimental data, which provides a measure of validation of the model.

%%%%%%%%%%%%%%%%%%%%%%%
\section{Discussion and concluding remarks}
%%%%%%%%%%%%%%%%%%%%%%%

The proposed dynamical model provides a rational basis for understanding the oncotripsy effect posited by Heyden and Ortiz \cite{Heyden2016} under the conditions of the experiments of Mittelstein {\sl et al}.~\cite{Mittelstein:2019}. An important difference between those experiments and the scenario initially contemplated in \cite{Heyden2016} is that in the experiments of Mittelstein {\sl et al}.~\cite{Mittelstein:2019} the cells are in aqueous suspension, whereas the analysis of Heyden and Ortiz \cite{Heyden2016} is concerned with cells embedded in a solid extracellular matrix (ECM). In aqueous suspension, the cells experience an exceedingly viscous environment, which is likely to suppress any vibrations of the cell membrane. The response of the cells to ultrasound stimulation is thus reduced to that of an internal resonator. Heyden and Ortiz \cite{Heyden2016} pointed out that the spectral gap between cancerous and healthy cells depends sensitively on the mechanical properties of the ECM and that the changes in those properties experienced by the cancerous tissue are a key contributing factor to the opening of a spectral gap. In addition, for cells embedded in an ECM, membrane rupture provides an additional lysis mechanism which is absent in cells in suspension. These considerations suggest the need for an independent experimental assessment of the oncotripsy effect in cancerous tissues, preferrably {\sl in vivo}.

The proposed dynamical model also reveals the dependence of oncotripsy on fundamental cell parameters and on process parameters. The cell parameters of the model can be calibrated from cell-death data for specific cell lines. Alternatively, fundamental cell properties such as stiffness and viscosity can be measured independently. The calibrated model can then be used as a tool for optimizing process parameters for maximum therapeutic effect. Most importantly, theoretical understanding such as provided by the proposed dynamical model is key for interpreting experimental observations and formulating new and improved clinical therapies.

In this regard, a number of possible therapies suggest themselves as possible clinical applications of oncotripsy. Thus, due to genomic instability and being in different states within the cell cycle, cancer cells are highly heterogeneous at any given moment. As such, it is unlikely that an entire cancer cell population can be killed by a single set of acoustic parameters. This suggests exploiting oncotripsy in connection with other synergistic cancer therapies such as immunogenic cell death (ICD). In this combination, oncotripsy does not need to kill every last cancer cell to be effective, as long as it can induce ICD of sufficient cancer cells to trigger the host immune system to kill remaining cancer cells (abscopal effect). Again, these and other fundamental questions suggest worthwhile directions for further research.

\section*{Acknowledgements}

The support of the California Institute of Technology through the Rothenberg Innovation Initiative and through the Caltech--City of Hope Biomedical Research Initiative is gratefully acknowledged.

\end{document}